\documentclass[journal,10pt]{IEEEtran}

\usepackage[usenames,dvipsnames]{xcolor}

\usepackage{cite}
\usepackage{blindtext}
\usepackage{graphicx}
\usepackage{epsfig}
\usepackage{amsmath,amssymb,bm,stmaryrd}
\usepackage{delarray}
\usepackage{stfloats}
\usepackage{float}
\usepackage{multirow,hhline}
\usepackage{algorithm}
\usepackage{algorithmic}
\usepackage{siunitx}
\DeclareSIUnit\bps{bps}
\DeclareSIUnit\Torr{Torr}
\DeclareSIUnit\torr{Torr}
\DeclareSIUnit\sample{Sa}

\usepackage{arydshln}
\usepackage{cuted}
\usepackage{balance}
\usepackage{verbatim} 
\usepackage{chronology-vert}
\usepackage{tabularx}
\usepackage{longtable}
\usepackage[utf8]{inputenc}
\usepackage{eurosym}

\DeclareUnicodeCharacter{20AC}{\EUR}

\usepackage[normalem]{ulem}
\usepackage{etoolbox}
\usepackage{verbatim}

\usepackage[flushleft]{threeparttable}

\usepackage{xcolor}
\newcommand\ytl[2]{
        \parbox[b]{8em}{\hfill{\color{cyan}\bfseries\sffamily #1}~$\cdots\cdots$~}\makebox[0pt][c]{$\bullet$}\vrule\quad \parbox[c]{4.5cm}{\vspace{7pt}\color{red!40!black!80}\raggedright\sffamily #2.\\[7pt]}\\[-3pt]}

\usepackage{tikz}
\usetikzlibrary{arrows,shapes,positioning,shadows,trees}

\tikzset{
  basic/.style  = {draw, text width=8cm, drop shadow, font=\sffamily, rectangle},
  root/.style   = {basic, rounded corners=2pt, thin, align=center,
                   fill=blue!40},
  level 2/.style = {basic, rounded corners=6pt, thin,align=center, fill=gray!60,
                   text width=12em},
  level 3/.style = {basic, thin, align=left, fill=pink!60, text width=8.5em}, 
 level 4/.style = {basic, thin, align=left, fill=pink!60, text width=6.5em} 
}

\begin{document}

\title{
 Terahertz Band: The Last Piece of RF Spectrum Puzzle for Communication Systems}


\author{Hadeel Elayan, Osama Amin, Basem Shihada,  Raed M. Shubair, and Mohamed-Slim Alouini~\IEEEmembership{}

\thanks{H. Elayan is with the Department of Electrical and Computer Engineering, University of Toronto, ON M5S Canada (e-mail: hadeel.mohammad@mail.utoronto.ca)}

\thanks{O. Amin, B. Shihada and M.S. Alouini  are   with   CEMSE   Division, King  Abdullah  University  of  Science  and  Technology  (KAUST),  Thuwal, Makkah   Province,   Saudi   Arabia. (e-mail: {osama.amin, basem.shihada,slim.alouini}@kaust.edu.sa)}

\thanks{ R. M. Shubair is  with the Research Laboratory of Electronics, Massachusetts Institute of Technology, Cambridge, MA 02139 USA (e-mail:  rshubair@mit.edu).}}

\maketitle

\begin{abstract}
Ultra-high bandwidth, negligible latency and seamless communication for  devices and applications are envisioned as major milestones that will revolutionize the way by which societies  create, distribute and consume information. The remarkable expansion of wireless data traffic that we are witnessing recently has advocated the investigation of suitable regimes in the radio spectrum to satisfy users' escalating requirements and allow the development and exploitation of both massive capacity and  massive  connectivity  of    heterogeneous  infrastructures. To this end, the Terahertz (THz) frequency band (0.1-10 THz) has received noticeable attention in the research community as an ideal choice for scenarios involving high-speed transmission. Particularly, with the evolution of technologies and devices, advancements in THz communication is bridging the gap between the millimeter wave (mmW) and optical frequency
ranges. Moreover,  the IEEE 802.15 suite of standards has been issued to shape regulatory frameworks that will enable innovation and provide a complete solution  that crosses between wired and wireless boundaries at 100 Gbps. Nonetheless, despite the expediting progress witnessed in  THz wireless research, the THz band is still considered one of the least probed frequency bands. As such, in this work, we present an up-to-date review paper  to analyze the fundamental elements  and mechanisms associated with the THz system architecture. THz generation methods are first addressed by highlighting the recent progress in the electronics, photonics as well as  plasmonics technology. To complement the devices, we  introduce the recent channel models available for indoor, outdoor as well as nanoscale propagation at THz band frequencies. A comprehensive comparison is then presented between the THz wireless communication and its other  contenders  by treating in depth the limitations associated with  each communication technology. In addition, several applications of THz wireless  communication are discussed taking into account the various length scales at which such applications occur. Further, as standardization is a fundamental aspect in regulating wireless communication systems, we highlight the  milestones  achieved regarding THz standardization activities. Finally, a future outlook is provided by presenting and envisaging several potential use cases and attempts to   guide  the deployment of the THz frequency band and mitigate the challenges related to high frequency transmission.  \\   
 
\end{abstract}
\section{Introduction}

\begin{figure}[!] 
  \centering
  \includegraphics[width=4.3in]{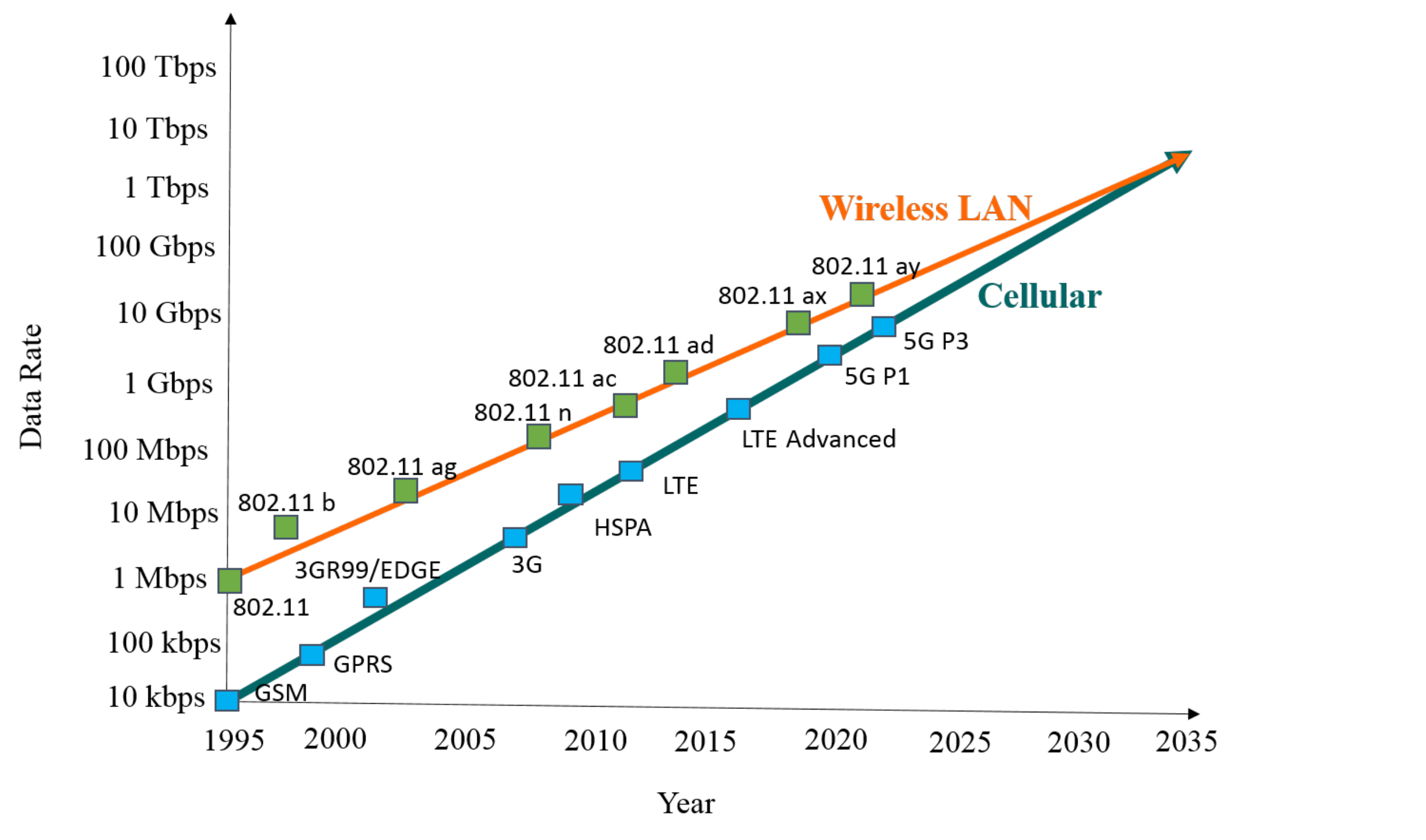}
  \caption{Wireless Roadmap  Outlook up to the year 2035.}
  \label{fig:wireless_cellular}
\end{figure}

  The race towards improving human life via developing different technologies  is witnessing a rapid pace  in diverse fields and at various scales. As for the integrated circuit field, the race focuses on increasing  the number of transistors on the wafer area, which is empirically predicted by Moore's Law  \cite{moore1998cramming}. In the case of the telecommunication sector, the race is moving towards boosting  the data rate to fulfill different growing service requirements, which is anticipated by Edholm's law of bandwidth  \cite{cherry2004edholm}. Wireless data traffic has been witnessing  unprecedented expansion in the past few years. On the one hand, mobile data traffic is anticipated to boost sevenfold between 2016 and 2021. On the other hand, video traffic is foreseeing a threefold increase during the same time period \cite{CISCO}. Actually, the traffic of both wireless and mobile devices is predicted to represent $ 71 \% $ of the total traffic by 2022 \cite{cisco2018cisco}.  In fact, by 2030, wireless data rates will be sufficient to compete with wired broadband \cite{li2018towards} as demonstrated in Fig. \ref{fig:wireless_cellular}. Such significant growth of wireless  usage has led the research community to explore appropriate regions in the radio spectrum to satisfy the escalating needs of individuals. To this end, the Terahertz (THz) frequency band (0.1-10 THz) started to gain noticeable attention within the global community.  Seamless data transfer, unlimited bandwidth, microsecond latency, and ultra-fast download are all features  of the THz technology that is anticipated to revolutionize the telecommunications landscape and alter the route through which people communicate and access information.

\begin{figure}[h!] 
        \centering
        \includegraphics[width=3.5 in]{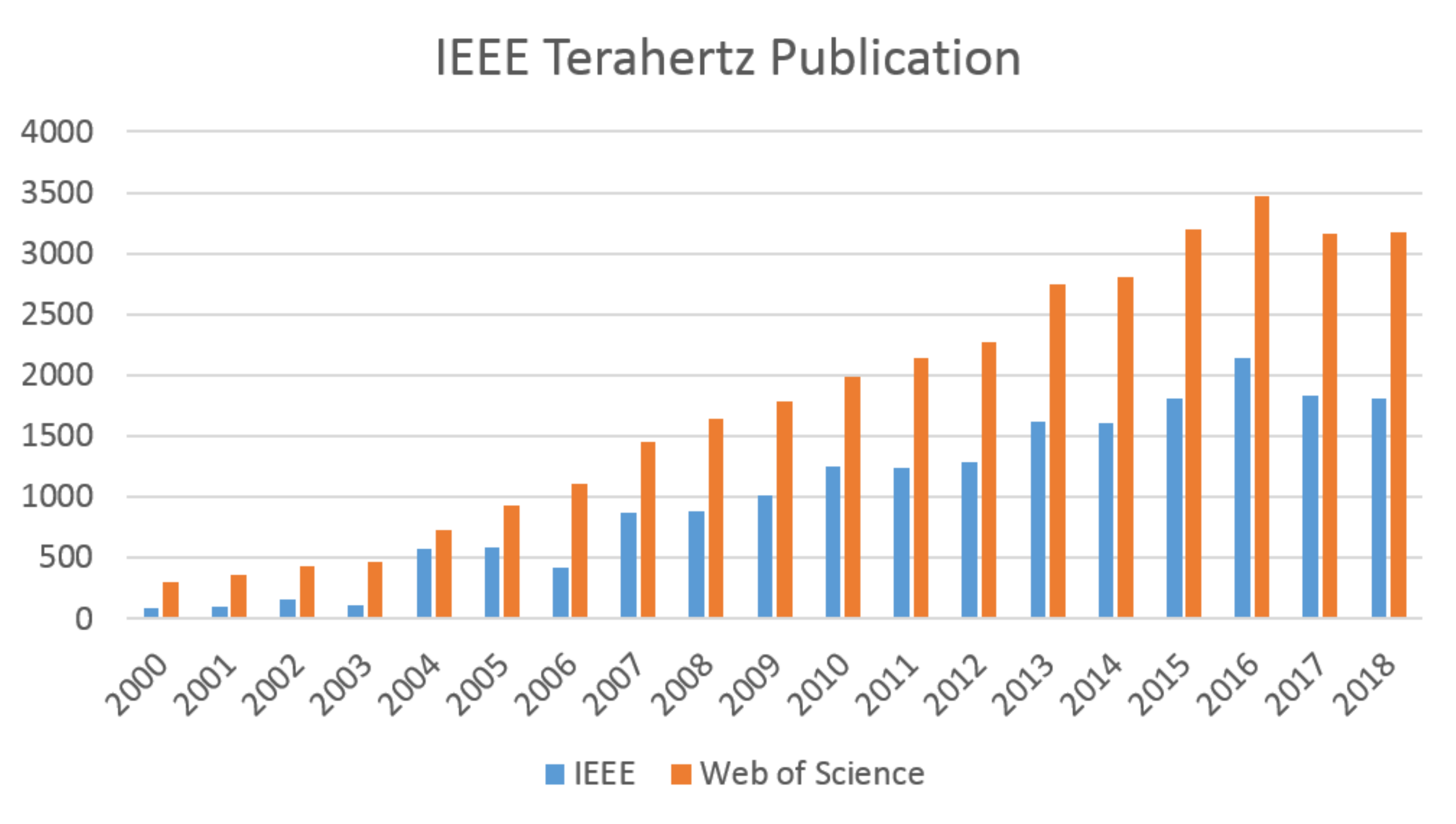}
        \caption{Terahertz publications  issued in both IEEE and web of science in recent years. }
        \label{fig:THz_pub}
  \end{figure}

 The  THz term has been first used within the microwave society during the 1970s to describe the spectral frequency of interferometers, diode detectors coverage, and water laser  resonance  \cite{kerecman1973tungsten, ashley1973transmission, fleming1974high}. During the 2000s, the THz term was referred to as the  submillimeter-wave with frequencies ranging between 100 GHz up to 10 THz. However, the boarder line between the submillimeter-waves and far infrared at that time was not clearly identified \cite{siegel2002terahertz, ferguson2002materials}. The concept of utilizing the THz for ultra-broadband communication using non-line of sight (NLoS) signal components  has been first proposed  as a powerful solution for extremely high data rates in \cite{piesiewicz2007short}. Since then,  THz technology in general and communication in particular grasped the enthusiasm of the research community. This interest has been reflected in the increased number of publications issued in both IEEE and web of science in recent years as demonstrated in Fig. \ref{fig:THz_pub}.

The THz frequency band  assures extensive throughput, which theoretically extends up to several THz leading to capacities in the order of Terabits per second (Tbps) \cite{6882305}. Such potential  associated with THz technology attracted the broader research community. In fact, the combined efforts of active research groups  is resulting in new designs, materials and fabrication methods that demonstrate   endless opportunities for THz  development.  Table~\ref{Table:THz_Groups} presents examples of  various groups that conduct THz  research, which indicated that research in this area is executed in laboratories across the globe.
Consequently, various funding agencies have been supporting THz projects and opening up new horizons in communications and devices deployed for beyond 5G technology. A detailed list of the most recent THz projects is demonstrated in Table~\ref{table:fund}.
 
 \begin{table}
        \caption{Research Groups working on THz Communication Related Topics}
        \label{Table:THz_Groups}
 
\begin{tabular}{ m{3cm}|m{1cm}|m{3.6cm} }  
        \footnotesize{\textbf{Research Group/Lab} }     & \footnotesize{\textbf{Location } }&\footnotesize{\textbf{ R$\&$D Activities }} \\ \hline               
                 
   \raisebox{10pt}
                    
       \footnotesize{Mittleman Lab at Brown University.} &  \footnotesize{USA}  & \footnotesize{THz  PHY layer, THz spectroscopy, THz probes.}
 \\ \hline

       \footnotesize{Broadband Wireless Networking Lab at Georgia Institute of Technology.} &  \footnotesize{USA}  & \footnotesize{THz PHY layer, THz MAC layer, THz Nanocommunication, THz devices.}  
         
     \\ \hline 
                
       \footnotesize{NaNoNetworking Center in Catalunya.} &  \footnotesize{Spain}  & \footnotesize{THz Nanocommunication}
 
  \\ \hline
       \footnotesize{Ultra-broadband Nano-Communication Laboratory at  University at Buffalo.} &  \footnotesize{USA}  & \footnotesize{THz PHY layer, THz MAC layer, THz Nanocommunication, THz devices.}    
 \\ \hline       
 
       \footnotesize{Terahertz Electronics Laboratory at UCLA.} &  \footnotesize{USA}  & \footnotesize{THz sources, detectors, spectrometers, reconfigurable meta-films, imaging and spectroscopy.}

 \\ \hline          
       \footnotesize{MIT Terahertz Integrated Electronics Group } &  \footnotesize{USA}  & \footnotesize{Sensing, metrology, security and communication at THz frequencies.}
 
  \\ \hline           
                \raisebox{10pt}
 
 \footnotesize{Fraunhofer Institute for Applied Solid State Physics IAF} & \footnotesize{Germany}  & \footnotesize{THz  PHY layer, MAC layer and RF electronics. } 

\\ \hline           
 
 \footnotesize{Terahertz Communications  Lab} & \footnotesize{Germany}  & \footnotesize{Channel investigation and                        Terahertz reflectors.} 

                         \\ \hline
               
                              \footnotesize{Core technology laboratory group in Nippon telegraph  and telephone (NTT) corporation}& \footnotesize{Japan}  & \footnotesize{Terahertz IC and Modularization Technology.}

                         \\ \hline
        
                \footnotesize{Texas Instrument Kilby  Lab} &  \footnotesize{USA}  & \footnotesize{ Ultra-Low Power Sub-THz CMOS Systems.}   
                
                  \\ \hline

                \footnotesize{Tonouchi Lab at Osaka university} &  \footnotesize{Japan}  & \footnotesize{ THz  Nanoscience, THz Bioscience,  THz-Bio sensing, and industrial applications.} 
               
                    \\ \hline
                        
                    \footnotesize{THz Electronics Systems Lab at Korea University.} &  \footnotesize{Korea}  & \footnotesize{ THz  PHY layer, MAC layer and RF electronics.} 
                    
           \\ \hline

                    \footnotesize{Nanocommunications Center at Tampere University of Technology.} &  \footnotesize{Finland}  & \footnotesize{ THz  PHY layer, THz  Nanocommunication.} 
                    
           \\ \hline
\end{tabular}
                          
\end{table}

\begin{table*}[!]
\caption{Examples of the Recent Funded THz Projects}
\begin{center}
\centering
\label{table:fund}
\small
\begin{tabular}{|p{130pt}| p{100pt}|p{40pt}|p{34pt}| p{35pt}|p{124pt}|}\hline
\raisebox{10pt}

\begin{footnotesize} \textbf{Project Title} \end{footnotesize}  & \begin{footnotesize}\textbf{Funding Agency}\end{footnotesize}& \begin{footnotesize} \textbf{Start Date}\end{footnotesize}& \begin{footnotesize} \textbf{End Date}\end{footnotesize} & \begin{footnotesize} \textbf{Fund}\end{footnotesize}&\begin{footnotesize} \textbf{Objective}\end{footnotesize} \\ \hline
\raisebox{10pt}

\begin{footnotesize}The Research and Development Project for Expansion of Radio Spectrum Resources. \end{footnotesize}&\begin{footnotesize}The Ministry of information and communications in Japan, and  the ministry of Education, Science, Sports and Culture.\end{footnotesize}& \begin{footnotesize}2008\end{footnotesize} & \begin{footnotesize}N/A\end{footnotesize} & \begin{footnotesize}N/A\end{footnotesize} & \begin{footnotesize}Developing technology for efficient frequency use, promoting shared frequency use, and encouraging a shift to use of higher frequencies. \end{footnotesize}\\ \hline
\raisebox{10pt}

\begin{footnotesize}Wireless Local Area Communication Systems at Terahertz Band.  \end{footnotesize} & \begin{footnotesize} Korea Government Funding Agency, IITA \end{footnotesize} & \begin{footnotesize}2008\end{footnotesize} & \begin{footnotesize}2012 \end{footnotesize}& \begin{footnotesize}25M \$\end{footnotesize} & \begin{footnotesize}Developing wireless LAN/PAN systems based on electronic devices.\end{footnotesize} \\ \hline
\raisebox{10pt}

\begin{footnotesize}Semiconductor Nanodevices for Room Temperature THz Emission and Detection (ROOTHz Project). \end{footnotesize} & \begin{footnotesize}  Framework Programmes for Research and Technological Development, European Union  \end{footnotesize} & \begin{footnotesize}2010\end{footnotesize} & \begin{footnotesize}2013\end{footnotesize} & \begin{footnotesize}2.1 M \EUR\  \end{footnotesize} & \begin{footnotesize}Fabricating solid state emitters and detectors at THz frequencies.\end{footnotesize}     \\ \hline
\raisebox{10pt}

\begin{footnotesize}TERAPAN: Ultra-high Data rate transmission with steerable antennas at THz Frequencies.\end{footnotesize}&\begin{footnotesize}German Federal Ministry of Education and Research\end{footnotesize}& \begin{footnotesize}2013\end{footnotesize} & \begin{footnotesize}2016\end{footnotesize} & \begin{footnotesize}1.5M \EUR\ \end{footnotesize} & \begin{footnotesize}Demonstrating adaptive wireless point-to-point THz communication for indoor environments at data rates of up to 100 Gbps.\end{footnotesize}  \\ \hline
\raisebox{10pt}

\begin{footnotesize}iBROW: Innovative ultra-BROadband ubiquitous Wireless communications through terahertz transceivers.  \end{footnotesize}&\begin{footnotesize}European Union’s Horizon 2020 research and innovation program \end{footnotesize}& \begin{footnotesize}2015\end{footnotesize} & \begin{footnotesize}2018\end{footnotesize} & \begin{footnotesize}4M  \EUR\ \end{footnotesize}& \begin{footnotesize}Developing novel, low cost, energy-efficient and compact ultra-broadband short-range wireless communication transceiver technology.  \end{footnotesize} \\ \hline
\raisebox{10pt}

\begin{footnotesize}TERAPOD: Terahertz based Ultra High Bandwidth Wireless Access Networks.\end{footnotesize}&\begin{footnotesize}European Union’s Horizon 2020 research and innovation program\end{footnotesize}& \begin{footnotesize}2017\end{footnotesize} & \begin{footnotesize}2020\end{footnotesize} & \begin{footnotesize}3.47M  \EUR\ \end{footnotesize} &\begin{footnotesize} Demonstrating  THz wireless link within a data centre proof of concept deployment as well as investigating other use cases  to beyond 5G. \end{footnotesize} \\ \hline
\raisebox{10pt}

\begin{footnotesize}ThoR: TeraHertz end-to-end wireless systems supporting ultra high data Rate applications.  \end{footnotesize}& \begin{footnotesize}European Union’s Horizon 2020 research and innovation program, and the National Institute of Information and Communications Technology in Japan (NICT)   \end{footnotesize}&\begin{footnotesize}2018\end{footnotesize}&\begin{footnotesize}2021\end{footnotesize} & \begin{footnotesize}1.5M  \EUR\ \end{footnotesize} & \begin{footnotesize}Providing technical solutions for the backhauling and fronthauling of  traffic at the  spectrum range near 300 GHz, which is able to cover  data rates required for beyond 5G systems. \end{footnotesize} \\ \hline
\raisebox{10pt}

\begin{footnotesize}ULTRAWAVE: Ultra capacity wireless layer beyond 100 GHz based on millimeter wave Traveling Wave Tubes.\end{footnotesize}& \begin{footnotesize}European Union’s Horizon 2020 research and innovation program \end{footnotesize}&\begin{footnotesize}2017\end{footnotesize}&\begin{footnotesize}2020\end{footnotesize} & \begin{footnotesize}3M  \EUR\ \end{footnotesize} & \begin{footnotesize}Developing a high capacity backhaul that enables 5G cell densification by exploiting bands beyond 100 GHz.\end{footnotesize} \\ \hline
\raisebox{10pt}

\begin{footnotesize}TERRANOVA: Terabit/s Wireless Connectivity by TeraHertz
innovative technologies to deliver Optical Network Quality of Experience in Systems beyond 5G. \end{footnotesize}& \begin{footnotesize}European Union’s Horizon 2020 research and innovation program\end{footnotesize}&\begin{footnotesize}2017\end{footnotesize}&\begin{footnotesize}2019\end{footnotesize} &  \begin{footnotesize}3M  \EUR\ \end{footnotesize} & \begin{footnotesize}Providing reliable connectivity of  high data rates and almost zero-latency in networks beyond 5G and extending the fiber optic systems  to wireless. \end{footnotesize}\\\hline
\raisebox{10pt}

 \begin{footnotesize}EPIC: Enabling Practical Wireless Tb/s Communications with Next Generation Channel Coding.  \end{footnotesize}& \begin{footnotesize}European Union’s Horizon 2020 research and innovation program \end{footnotesize}& \begin{footnotesize}2017\end{footnotesize}& \begin{footnotesize}2020\end{footnotesize} & \begin{footnotesize}3M  \EUR\ \end{footnotesize} &  \begin{footnotesize}Developing new FEC codes to serve as an enabler of practicable beyond 5G wireless Tbps solutions. \end{footnotesize}\\\hline
  \raisebox{10pt}

\begin{footnotesize}DREAM: D-Band Radio solution Enabling up to 100 Gbps reconfigurable Approach for Meshed beyond 5G networks.  \end{footnotesize}& \begin{footnotesize}European Union’s Horizon 2020 research and innovation program \end{footnotesize}& \begin{footnotesize}2017\end{footnotesize}& \begin{footnotesize}2020\end{footnotesize}& \begin{footnotesize}2.8M  \EUR\  \end{footnotesize}&\begin{footnotesize}Enabling wireless links with data rate exceeding current V-band and E-band backhaul solutions to bring wireless systems to the speed of optical systems.\end{footnotesize}\\\hline
   \raisebox{10pt}

 \begin{footnotesize}WORTECS: Wireless Optical/Radio TErabit Communications. \end{footnotesize}& \begin{footnotesize}European Union’s Horizon 2020 research and innovation program\end{footnotesize}& \begin{footnotesize}2017\end{footnotesize}& \begin{footnotesize}2020\end{footnotesize}& \begin{footnotesize}3M  \EUR\ \end{footnotesize} & \begin{footnotesize}Exploring Tbps capability of above 90 GHz spectrum, while combining radio and optical wireless technologies.\end{footnotesize}\\ \hline
  \raisebox{10pt}

 \begin{footnotesize}TerraNova: An Integrated Testbed for True Terahertz Communications.
 \end{footnotesize}& \begin{footnotesize}National Science Foundation (NSF)\end{footnotesize}& \begin{footnotesize}2017\end{footnotesize}& \begin{footnotesize}2019\end{footnotesize}& \begin{footnotesize}750K \$ \end{footnotesize}& \begin{footnotesize}Developing the first integrated testbed specific to ultra-broadband communication networks at  THz frequencies.\end{footnotesize}
 \\ \hline
 \raisebox{10pt}

 \begin{footnotesize}EAGER: High-performance Optical-phonon-based Terahertz Sources Operating at Room Temperature.\end{footnotesize}& \begin{footnotesize}National Science Foundation (NSF)\end{footnotesize}& \begin{footnotesize}2017\end{footnotesize}& \begin{footnotesize}2018\end{footnotesize}&   \begin{footnotesize}85K \$ \end{footnotesize}& \begin{footnotesize} Systematically exploring how to realize a new type of THz sources based on fundamentally different device operation principles. \end{footnotesize}\\ \hline
 \raisebox{10pt}

 \begin{footnotesize}Novel Terahertz Generators Based on magnetic Materials.\end{footnotesize}& \begin{footnotesize}National Science Foundation (NSF)\end{footnotesize}& \begin{footnotesize}2017\end{footnotesize}& \begin{footnotesize}2020\end{footnotesize}& \begin{footnotesize}210K \$\end{footnotesize} &  \begin{footnotesize}Creating a new type of THz generators that are compact, inexpensive, and operate at room temperature by converting magnetic oscillations into THz  waves\end{footnotesize}. \\\hline

\end{tabular}
\end{center}
\end{table*}Several studies available in the literature reviewed and discussed the potential benefits that can be reaped from the THz band \cite{siegel2002terahertz}. The first THz survey was introduced in 2002 by Siegel and focused on the sources, sensors and applications for frequencies higher than 500 GHz \cite{siegel2002terahertz, ferguson2002materials, siegel2004terahertz, fitch2004terahertz}. During the same time period, another article has been issued in an attempt to demonstrate THz material characterization, which results in several applications  including THz imaging and tomography \cite{ferguson2002materials}. From a medical and biological perspective, Siegel reviewed in \cite{siegel2004terahertz} the developments observed in THz irradiation and sensing. In \cite{fitch2004terahertz}, Fitch and Osiander presented the first overview of THz technology for various practical deployments in communications and sensing including security and spectroscopy applications. After that, the promise brought by THz frequencies ranging from 100 GHz up to 30 THz has been demonstrated in \cite{tonouchi2007cutting}, where   discussions in terms of generation techniques and their correlated output power abilities have been presented. In \cite{jacob2009overview}, Jacob \textit{et al.} provided a brief overview of the research activities  including channel modeling and signal generation in both the millimeter wave   (mmW) and THz bands. The first review on THz communication systems was presented in 2010, where Federici and Moeller presented a focused discussion on channel model basic considerations, THz generation methods and implementation issues of THz communications \cite{federici2010review}. In \cite{kleine2011review}, Kleine-Ostmann and Tadao Nagatsuma  further expanded the discussion on the research progress in THz technology. In \cite{song2011present}, Song and Nagatsuma shed the light on some advances  of THz communication including achievable  data rates and service distances in addition to highlighting the challenges associated with the 275 GHz up to 3 THz frequency band. A similar and brief review has been introduced by  Nagatsuma in  \cite{nagatsuma2011terahertz}, which focused on demonstrations from 100 GHz to 300 GHz. In \cite{huang2011terahertz}, Huang \textit{et al.} provided  both an  overview of  the  state-of-the-art in  THz  wireless  communication  along with  a tutorial for  emerging applications in Terabit radio systems. In \cite{nagatsuma2013terahertz}, Nagatsuma \textit{et al.} reviewed the progress in photonics technology in generating THz signals  ranging from 100 GHz to 300 GHz. In \cite{akyildiz2014terahertz}, Akyildiz \textit{et al.} summarized the THz possible applications in wireless communications and defined the challenges of this promising band. In \cite{kurner2014towards}, K\"{u}rner and  Priebe demonstrated more applications and reviewed briefly some research in THz communication.  In \cite{hirata2015ultrafast}, Hirata and Yaita discussed several THz technologies related to devices, circuits and antennas  in addition to some recent experimental test-beds. In \cite{petrov2016terahertz}, Petrov \textit{et al.} discussed further applications and defined major research challenges besides showcasing  the progress towards THz standardization. In \cite{7982949}, Mumtaz  \textit{et al.} overviewed the opportunities and challenges in THz communications for vehicular networks indicating that communication at much higher frequencies  is correlated with considerable potential when it comes to vehicular networks. In \cite{mittleman2017perspective}, Mittleman presented a perspective article where he highlighted several breakthroughs in the THz field which enabled  new opportunities for both fundamental and applied research.  The author emphasized on how the achievements of integrated THz sources and systems continue to accelerate enabling many new applications.   In \cite{sengupta2018terahertz}, Sengupta \textit{et al.} reviewed the current progress in generating THz signals using electronics and hybrid electronics-photonics systems for communication, sensing and imaging applications. Recently, in\cite{chen2019survey},  Chen \textit{et al.} provided a  literature review on the development towards THz communications and presented key technical challenges faced in THz wireless communication systems. In \cite{ghafoor2019mac},  from the Medium Access Protocol (MAC) perspective, Ghafoor \textit{et al.} presented an in-depth survey of THz MAC protocols highlighting key features which should be considered while designing efficient protocols. In \cite{tekbiyik2019terahertz}, Tekbiyik \textit{et al.} addressed the current open issues in the design of THz wireless communication systems in terms of hardware, physical channel and network. Finally, in \cite{Rappaport2019}, Rappaport \textit{et al.} presented a number of promising  approaches and novel approaches that will aid in the development and implementation of the sixth generation (6G) of wireless networks using THz frequencies. The aforementioned review articles are listed in  Table \ref{Table:Table_Surveys}, indicating clearly  a high activity rate since the early time of 2000 as a result of the advances in both electronic and photonic technologies and the demand to fulfill several application requirements.   To this end, there is still a demand to have a comprehensive view on the current progress and recent advances in this field that would help researchers  draw futuristic steps for several communication systems. As such, this paper  aims to   serve such an objective by presenting the latest technologies associated with the THz frequency band.

 \begin{table*}[!] 
        \caption{Terahertz Technology Surveys in the Literature}
        \begin{center}
                \centering
                \tiny
               \label{Table:Table_Surveys}
                \begin{tabular}{|p{4pt}| p{180pt}|p{50pt}|p{165pt}|p{30pt}|}\hline
                        
                        \raisebox{10pt}
                        
                        \begin{footnotesize}\textbf{}\end{footnotesize}&\begin{footnotesize} Survey Title\end{footnotesize} & \begin{footnotesize}Year Published\end{footnotesize}& \begin{footnotesize}Survey Content\end{footnotesize}&\begin{footnotesize}Reference\end{footnotesize}\\\hline
                        
                        \raisebox{10pt}
                        
                        \begin{footnotesize}\textbf{1}\end{footnotesize}&\begin{footnotesize} Terahertz Technology\end{footnotesize} & \begin{footnotesize}2002\end{footnotesize} & \begin{footnotesize} The first review article on the applications, sources and sensors for the THz technology with the emphasis on frequencies higher than 500 GHz.
                        \end{footnotesize}& \begin{footnotesize}\cite{siegel2002terahertz}\end{footnotesize}\\\hline
                        
                        \raisebox{10pt}
                        
                        \begin{footnotesize}\textbf{2}\end{footnotesize}&\begin{footnotesize}  Materials for terahertz science and technology \end{footnotesize} & \begin{footnotesize}2002\end{footnotesize} & \begin{footnotesize} The article presents a review on material research in developing THz sources and  detectors to support different applications.
                        \end{footnotesize}& \begin{footnotesize}\cite{ferguson2002materials}\end{footnotesize}\\\hline
                        
                        \raisebox{10pt}
                        
                        \begin{footnotesize}\textbf{3}\end{footnotesize} &\begin{footnotesize}   Technology in Biology and Medicine \end{footnotesize}  &\begin{footnotesize} 2004 \end{footnotesize}&\begin{footnotesize} The emerging field of THz   is surveyed in biology and medicine, in which the irradiation and sensing capabilities of THz waves are applied for different applications.  
                        \end{footnotesize}& \begin{footnotesize} \cite{siegel2004terahertz} \end{footnotesize}\\\hline
                        
                        \raisebox{10pt}
                        
                        \begin{footnotesize}\textbf{4}\end{footnotesize} &\begin{footnotesize} Terahertz waves for communications and sensing\end{footnotesize}  &\begin{footnotesize} 2004 \end{footnotesize}&\begin{footnotesize} This survey gives an overview of THz technology in terms of sources, detectors, and modulators needed for several applications such as security and spectroscopy.  
                        \end{footnotesize}& \begin{footnotesize}\cite{fitch2004terahertz}\end{footnotesize}\\\hline
                        
                        \raisebox{10pt}
                        
                        \begin{footnotesize}\textbf{5}\end{footnotesize} &\begin{footnotesize} Cutting-edge terahertz technology \end{footnotesize}  &\begin{footnotesize} 2007 \end{footnotesize}&\begin{footnotesize}This review article gives   
                                an overview of the THz technology progress status and expected usages in wireless communication, agriculture and medical applications.
                        \end{footnotesize}& \begin{footnotesize}\cite{tonouchi2007cutting}\end{footnotesize}\\\hline
                        
                        \raisebox{10pt}
                        
                        \begin{footnotesize}\textbf{6}\end{footnotesize} &\begin{footnotesize} An Overview of Ongoing Activities in the Field of Channel Modeling, Spectrum Allocation and Standardization for mm-Wave and THz Indoor Communications \end{footnotesize}  &\begin{footnotesize} 2009 \end{footnotesize}&\begin{footnotesize} An overview of mm-Wave and THz radio  channel modeling along with some  investigation results are presented. The article also discusses the    status of standardization activities and plans.
                        \end{footnotesize}& \begin{footnotesize}\cite{jacob2009overview}\end{footnotesize}\\\hline
                        
                        \raisebox{10pt}
                        
                        \begin{footnotesize}\textbf{7}\end{footnotesize} &\begin{footnotesize} Review of Terahertz and Subterahertz Wireless Communications\end{footnotesize}  &\begin{footnotesize} 2010\end{footnotesize}&\begin{footnotesize} 
                                The first review article on THz communication systems, which demonstrates basic channel modeling, generation methods,  detection, antennas, and a  summary of THz communication link measurements. 
                        \end{footnotesize}& \begin{footnotesize}\cite{federici2010review}\end{footnotesize}\\\hline
                        
                        \raisebox{10pt}
                        
                        \begin{footnotesize}\textbf{8} \end{footnotesize} &\begin{footnotesize} A Review on Terahertz Communications Research\end{footnotesize} &\begin{footnotesize} 2011\end{footnotesize}& \begin{footnotesize} A brief overview of emerging THz technologies, THz modulators, channel modeling and system research that might lead to future communication systems. \end{footnotesize} & \begin{footnotesize}\cite{kleine2011review}\end{footnotesize}\\\hline
                        
                        \raisebox{10pt}
                        
                        \begin{footnotesize}\textbf{9}\end{footnotesize} &\begin{footnotesize} Present and Future of Terahertz Communications \end{footnotesize} &\begin{footnotesize} 2011 \end{footnotesize}& \begin{footnotesize} A review on THz communication as  an alternative solution for high data rate future wireless communication systems, especially  short range networks.   \end{footnotesize}& \begin{footnotesize}\cite{song2011present}\end{footnotesize}\\\hline
                        
                        \raisebox{10pt}
                        
                        \begin{footnotesize}\textbf{10}\end{footnotesize} &\begin{footnotesize} Terahertz technologies: present and future \end{footnotesize} &\begin{footnotesize} 2011 \end{footnotesize}& \begin{footnotesize}  This paper overviews the  progress in THz technology and applications as well as summarizes the recent demonstrations from 100 GHz to 300 GHz.  \end{footnotesize}& \begin{footnotesize}\cite{nagatsuma2011terahertz}\end{footnotesize}\\\hline

 \raisebox{10pt}
                    
                    \begin{footnotesize}\textbf{11}\end{footnotesize} &\begin{footnotesize} Terahertz Terabit Wireless Communication \end{footnotesize} &\begin{footnotesize} 2011 \end{footnotesize}& \begin{footnotesize}The  state-of-the-art  in  THz  wireless  communication  along with the emerging applications in Terabit  radio systems are demonstrated. \end{footnotesize}& \begin{footnotesize}\cite{huang2011terahertz}\end{footnotesize}\\\hline

                        \raisebox{10pt}
                        
                        \begin{footnotesize}\textbf{12}\end{footnotesize} &\begin{footnotesize} Terahertz wireless communications based on photonics technologies \end{footnotesize} &\begin{footnotesize} 2013 \end{footnotesize}& \begin{footnotesize}This paper overviews the recent advances in THz generation using phonetics towards achieving up to 100 Gbps data rate either on real time or offline.    \end{footnotesize}& \begin{footnotesize}\cite{nagatsuma2013terahertz}\end{footnotesize}\\\hline
                        
                        \raisebox{10pt}
                        
                        \begin{footnotesize}\textbf{13}\end{footnotesize} &\begin{footnotesize}Terahertz band: Next frontier for Wireless Communications \end{footnotesize} &\begin{footnotesize} 2014  \end{footnotesize}& \begin{footnotesize} A review of THz applications and challenges in generation, channel modeling and communication systems is presented along with  a brief discussion on experimental and simulation testbeds.
                        \end{footnotesize}&  \begin{footnotesize}\cite{akyildiz2014terahertz}\end{footnotesize}\\\hline         
                        
                        \raisebox{10pt}
                        
                        \begin{footnotesize}\textbf{14}\end{footnotesize} &\begin{footnotesize} Towards THz Communications-status in research, standardization and regulation \end{footnotesize} &\begin{footnotesize} 2014  \end{footnotesize}& \begin{footnotesize}The article provides an overview of THz communications, research projects, spectrum regulations and ongoing standardization activities.\end{footnotesize}& \begin{footnotesize} \cite{kurner2014towards} \end{footnotesize}\\\hline
                        
                        \raisebox{10pt}
                        
                        \begin{footnotesize}\textbf{15}\end{footnotesize} &\begin{footnotesize}   Ultrafast terahertz wireless communications technologies \end{footnotesize} &\begin{footnotesize} 2015  \end{footnotesize}& \begin{footnotesize} The article provides an overview of THz communication research, development and implementation testbeds. \end{footnotesize}& \begin{footnotesize} \cite{hirata2015ultrafast} \end{footnotesize}\\\hline

                        \raisebox{10pt}
                        
                        \begin{footnotesize}\textbf{16}\end{footnotesize} &\begin{footnotesize} Terahertz Band Communications: Applications, Research Challenges, and Standardization Activities\end{footnotesize} &\begin{footnotesize}  2016 \end{footnotesize}& \begin{footnotesize} The article summarizes the recent achievements by industry, academia and standardization bodies  in  the THz field as well as discusses the open research challenges. \end{footnotesize}& \begin{footnotesize} \cite{petrov2016terahertz}\end{footnotesize}\\ \hline

 \raisebox{10pt}
                        
                                             \begin{footnotesize}\textbf{17}\end{footnotesize} &\begin{footnotesize} Terahertz Communication for Vehicular Networks \end{footnotesize} &\begin{footnotesize} 2017 \end{footnotesize}& \begin{footnotesize}An overview of the opportunities and challenges in THz communications for vehicular networks is provided.\end{footnotesize}& \begin{footnotesize}\cite{7982949}\end{footnotesize}

\\\hline

 \raisebox{10pt}
                        
                                             \begin{footnotesize}\textbf{18}\end{footnotesize} &\begin{footnotesize} Perspective: Terahertz science and technology \end{footnotesize} &\begin{footnotesize} 2017 \end{footnotesize}& \begin{footnotesize} The article discusses several breakthroughs in the THz field which enabled  new opportunities for both fundamental and applied research. \end{footnotesize}& \begin{footnotesize}\cite{mittleman2017perspective}\end{footnotesize}

\\\hline

 \raisebox{10pt}
                        
                        \begin{footnotesize}\textbf{19}\end{footnotesize} &\begin{footnotesize}  Terahertz integrated electronic and hybrid electronic–photonic systems \end{footnotesize} &\begin{footnotesize}  2018 \end{footnotesize}& \begin{footnotesize} The article reviews the development of THz integrated electronic and hybrid electronic–photonic systems used in several applications. \end{footnotesize}& \begin{footnotesize} \cite{sengupta2018terahertz}\end{footnotesize}\\ \hline

\raisebox{10pt}       

  \begin{footnotesize}\textbf{20}\end{footnotesize} &\begin{footnotesize} A Survey on Terahertz Communications \end{footnotesize} &\begin{footnotesize} 2019 \end{footnotesize}& \begin{footnotesize}The paper  provides
a  literature review on the development towards THz communications and presents some key technologies faced in THz wireless communication systems. \end{footnotesize}& \begin{footnotesize}\cite{chen2019survey}\end{footnotesize}\\\hline

\raisebox{10pt}

\begin{footnotesize}\textbf{21}\end{footnotesize} &\begin{footnotesize} MAC Protocols for Terahertz Communication:  
A
 Comprehensive Survey\end{footnotesize} &\begin{footnotesize} 2019 \end{footnotesize}& \begin{footnotesize}In this survey, detailed work on existing THz MAC
protocols with classifications, band features, design issues and
challenges are discussed.\end{footnotesize}& \begin{footnotesize}\cite{ghafoor2019mac}\end{footnotesize}
\\\hline

\raisebox{10pt}       

  \begin{footnotesize}\textbf{22}\end{footnotesize} &\begin{footnotesize} Terahertz band communication systems: Challenges, novelties and standardization efforts \end{footnotesize} &\begin{footnotesize} 2019 \end{footnotesize}& \begin{footnotesize}The paper addresses the current open issues in the design of THz wireless communication system in terms of hardware, physical channel and network.\end{footnotesize}& \begin{footnotesize}\cite{tekbiyik2019terahertz}\end{footnotesize}\\\hline 
\raisebox{10pt} 

\begin{footnotesize}\textbf{23}\end{footnotesize} &\begin{footnotesize} Wireless Communications and Applications Above
100 GHz: Opportunities and Challenges for 6G and
Beyond\end{footnotesize} &\begin{footnotesize} 2019 \end{footnotesize}& \begin{footnotesize}The paper presents a number of promising  approaches  that will aid in the development and implementation of the 6G wireless networks.\end{footnotesize}& \begin{footnotesize}\cite{Rappaport2019}\end{footnotesize}\\\hline
                        %
                        %
                \end{tabular}
        \end{center}
 \end{table*}

Due to the rise of wireless traffic,  the interest in higher bandwidth will never seem  to descend before the capacity of  the technology even beyond 5G  has attained an upper bound \cite{elayan2018terahertz}. In this paper, we shed the light on various opportunities associated  with the deployment of the THz frequency band. These opportunities are  demonstrated as  applications that will facilitate a refined wireless experience coping  with users' needs. Therefore, the main objective of the presented work is to provide the reader with an in-depth discussion, in which the authors summarize the latest literature findings regarding the fundamental aspects of THz frequency band wireless communication. The presented work will help researchers determine the gaps available in the literature paving the way for the research community to further develop research in the field.  The rest of the paper is organized as follows. In Section~\ref{Sec2},  we  review the THz frequency band   generation techniques  available in the literature. In Section~\ref{Sec3}, the THz channel models which capture the channel characteristics  and propagation phenomena are presented. In Section~\ref{Sec4}, an extensive comparison  is conducted in order to highlight the differences between  THz wireless and other existing technologies including  mmW, infrared, visible light and ultraviolet communication. In Section~\ref{Sec5}, diverse applications which tackle nano, micro as well as macro-scale THz scenarios are presented. In Section~\ref{Sec6}, the standardization activities involved in regulating the usage of THz communication are extensively discussed. In Section~\ref{Sec7}, a plethora of opportunities brought by the deployment of the THz frequency band are demonstrated in an aim to effectively meet the needs of future networks and face the technical challenges associated with implementing THz communication. Finally, we conclude the paper in Section~\ref{Sec8}.


\section{Terahertz Frequency Generation Methods}
\label{Sec2}
\begin{figure}[h!]
        \centering
\input{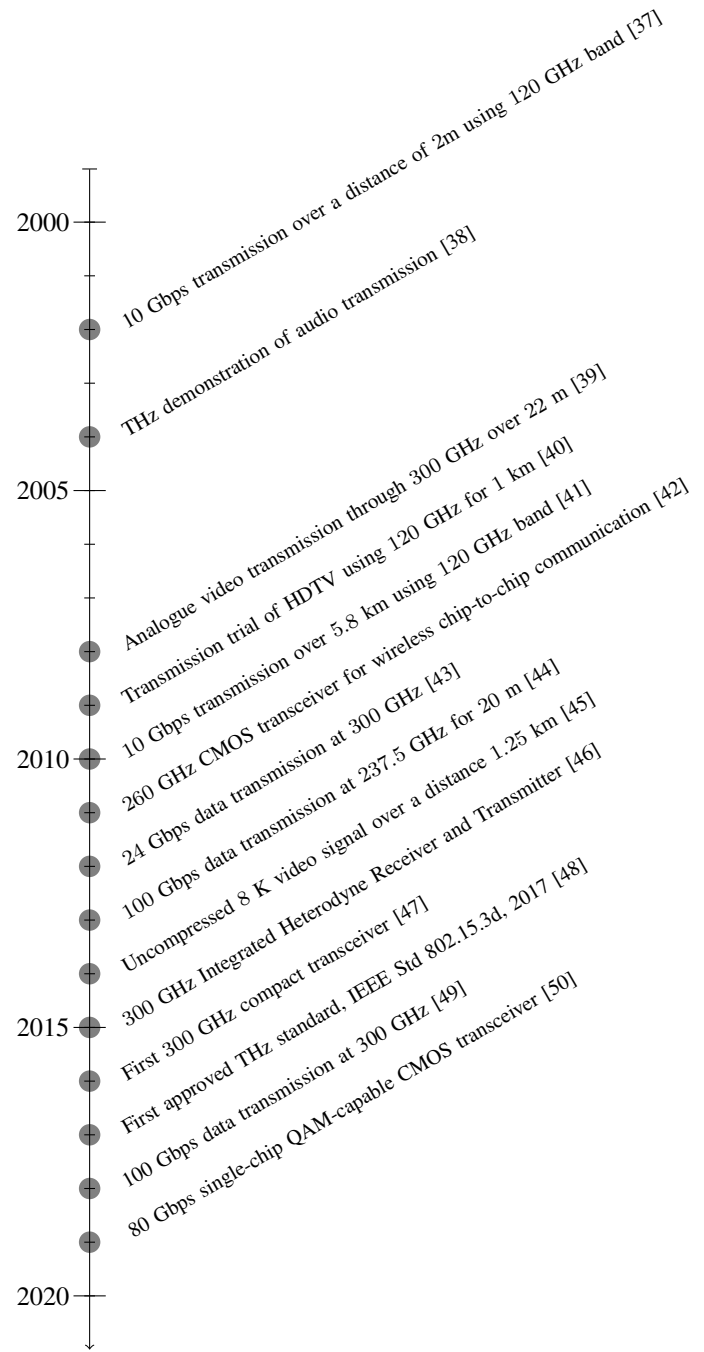}
\caption{Time-line of Progress in Terahertz Communication Technology.}
\label{fig:timeline}
\end{figure}

In recent years, broadband wireless links using the THz frequency band have been attracting the interests of  research groups worldwide.  By utilizing the frequency range above 100 GHz, the potential to employ extremely large bandwidths  and achieve data rates exceeding 100 Gbps for radio communications will eventually be enabled. Nevertheless, in order to fulfill such aim, progress from the devices perspective is a necessity. In fact, the location of the THz band between the microwave and infrared frequency ranges imposes difficulty on signal generation and detection. Therefore, the frequency range between 0.1 and 10 THz has been often referred to as the THz Gap since the technologies used for generating and detecting such radiation is considered less mature. On the one hand, transistors and other quantum devices which rely on electron transport are limited to about 300 GHz. Devices functioning above these frequencies tend to be  inefficient as semiconductor technologies fail to effectively convert electrical power into electromagnetic radiation at such range \cite{zhu2011ultrafast}.  Operating at high frequencies  requires rapidly alternating currents; thus,  electrons  will not be capable of travelling far enough to enable a device to work before the polarity of the voltage changes and the electrons change direction. On the other hand, the wavelength of photonic devices can be extended down to only 10 $\mu$m (about 30 THz). This is due to the fact that electrons   move vigorously  between energy levels resulting in a difficulty to control the small discrete energy jumps needed to release photons with THz frequencies. Hence, designing optical systems with dimensions close to THz wavelengths is a challenge \cite{lee2007searching}. Nonetheless, with the development of novel techniques, often combining electronics and photonics approaches, THz research is recently being pushed into the center stage.   Fig.~\ref{fig:timeline} presents  a time-line of the progress in THz communication technology indicating how THz research is moving from an emerging to a more established field, where  an obvious  technological leap has been witnessed within the last decade. The following subsections discuss the latest THz advancements  achieved focusing mainly on both the electronics  and photonics fields while shedding the light on other techniques used to generate THz waves. In particular, Table~\ref{table:Advances} summarizes the advancements in THz technology by presenting the progress  over the years in THz electronic as well as photonic transceivers, achievable data rates and propagating distances as well as output power. 

\subsection{Solid-State Electronics}
 Recent advances in the development of semiconductor components and their manufacturing technology are making THz systems both feasible and affordable resulting in compact devices. In fact, technology  limitations  have been overcome  by  architectural  innovations as  well  as  by  new device  structures.

\subsubsection{Complementary Metal-Oxide-Semiconductor (CMOS)} 

CMOS-based sources have been developing rapidly in recent years.  Such technology possesses the advantages of high level integration, small form factor, and potential low cost. The high frequency operation ability of CMOS offers solutions in the lower band of the THz spectrum. This has been achieved by adding either a Voltage Controlled Oscillator (VCO) or inserting an active multiplier chain in the CMOS device \cite{7123677}. Various triplers are used to multiply the frequency from a lower band to the THz frequency band by using nanoscale CMOS technology, where the consideration for CMOS THz circuits is enabled by   technology  scaling.  In 2006, the scaling of a 65-nm CMOS process has resulted in  a power gain frequency of 420 GHz, in  which uni-axial strained silicon transistors with physical gate lengths of 29-nm have been used \cite{post200665nm}. In 2007,  a transistor cutoff frequency of 485 GHz \cite{lee2007record} has been achieved while utilizing a 45-nm microprocessor technology.  The authors in \cite{shim2011553} demonstrated a 553 GHz quadruple-push oscillator using 45-nm CMOS technology, while in \cite{6814332} the authors presented a 540 GHz signal generator fabricated in 40-nm  bulk CMOS. In addition, the authors in \cite{zhao20160} presented a 560 GHz frequency synthesizer realized in 65-nm CMOS technology. The chip configuration constituted of both a THz VCO along with a phase locked loop circuit. As such, it could be noticed that the constructive addition of harmonic signals allows  devices to penetrate into hundreds of GHz range which indicates the impending THz era of CMOS technology. Such results states that the industry has been capable of keeping up with the documents reported by the International Roadmap for Semiconductors \cite{wilson2013international}. CMOS  transmitters have actually achieved up to  105 Gbps data rate using a 40-nm CMOS process  at 300 GHz \cite{7870384}.

\subsubsection{Monolithic Microwave Integrated Circuits (MMIC)}
Assimilating a large number of tiny transistors into a small chip leads  to circuits that are orders of magnitude smaller, cheaper, and faster than those built of discrete electronic components. Critical for reaching THz operational frequencies for integrated circuits are transistors with sufficiently high maximum  oscillation frequency, $f_{max}$. The main approaches in developing high speed transistors include both transistor gate  scaling for parasitic reduction as well  as  epitaxial  material enhancement for improved electron transport properties. A variety of MMIC compatible processes include  Heterojunction Bipolar Transistors (HBTs) and High Electron Mobility Transistors (HEMT). Both transistors use different semiconductor materials for the emitter and base regions, creating a heterojunction which limits the injection of holes from the base into the emitter. This allows  high doping density to be used in the base which results in reducing the base resistance while maintaining gain. In comparison to conventional bipolar transistors, HBTs have the advantage of higher cut-off frequency, higher voltage handling capability and reduced capacitive coupling  with the substrate \cite{880842}. Materials used for the substrate include silicon, gallium arsenide (GaAs), and indium phosphide (InP). Both  GaAs and InP  HBTs are compatible for integration with 1.3-1.5 $\mu$m optoelectronics such as lasers and photodetectors. In the case of HEMTs,  the most commonly used material combination in the literature involves GaAs. Nonetheless, gallium nitride (GaN) HEMTs in recent years have attracted attention due to their high-power performance.  GaN HEMT technology is promising for broadband
wireless communication systems because of its high breakdown electric field and high saturation carrier velocity compared to other competing technologies such as GaAs and InP devices \cite{masuda2009gan}. In fact, by utilizing a MMIC GaAs HEMT front-end, data rates up to 64 Gbps over 850 m \cite{kallfass201564} and 96 Gbps over 6 m \cite{6956202}  have been attained using a 240 GHz carrier frequency. In terms of InP-HEMT, improvement in electron-beam lithography is witnessing the increase in the speed  of  such devices  as gate length decreases. A significant milestone was the first InP HEMT with $f_{\max}$ $>$ 1 THz reported in 2007 \cite{4419013}. Further milestone achievements in amplifications at higher frequencies have been demonstrated with subsequent generation of transistors and designs at 480 GHz \cite{deal2010demonstration}, 670 GHz \cite{deal2011low}, and 850 GHz \cite{deal2014recent}. By using 25-nm gate InP  HEMT, $f_{\max}$ reached 1.5 THz \cite{mei2015first}. Several devices with high $f_{\max}$ that operate around 1 THz are reported in Table \ref{table:HEMT-technology}. 

Compared with CMOS, higher frequency sources with higher output powers have been obtained in the literature using HBT and HEMT technologies \cite{etsi2016mmwave}. Nonetheless, CMOS still remains an attractive candidate for THz technology due to its lower cost and higher integration densities. It is to be noted  that the development of physical principles of THz-wave amplification and oscillation is one of problems hindering progress in modern solid state electronics towards high frequencies. Therefore, novel perspectives are tied with use of resonant tunneling quantum effects, characterized by short transient times in comparison to the fast response of superconducting devices as will be discussed in the subsequent section.
\begin{table}[h!]
\caption{Progress of InP HEMT  in relation to  oscillation frequency and gate length}
\begin{center}
\centering
\label{table:HEMT-technology}
\small
\begin{tabular}{|p{45pt}| p{45pt}|p{34pt}|}\hline

\raisebox{10pt}

\begin{footnotesize} \textbf{Gate Length} \end{footnotesize}  &
 \begin{footnotesize} $ \mathbf{f_{\max}}  (\mathbf{THz})$  \end{footnotesize} & \begin{footnotesize} \textbf{Reference}\end{footnotesize} \\ \hline

\raisebox{10pt}

\begin{footnotesize}75 nm\end{footnotesize}  &\begin{footnotesize} 0.91  \end{footnotesize}& \cite{7747517}\\\hline
\raisebox{10pt}

\begin{footnotesize}{75 nm}\end{footnotesize}  &\begin{footnotesize}1.3  \end{footnotesize}& \cite{takahashi2017maximum}\\\hline

\raisebox{10pt}

\begin{footnotesize}{50 nm}\end{footnotesize} &\begin{footnotesize}  1.1 \end{footnotesize}& \cite{4419013}\\\hline

\raisebox{10pt}

\begin{footnotesize}50 nm\end{footnotesize} & \begin{footnotesize} 1.06 \end{footnotesize}& \cite{5703453}\\\hline

\raisebox{10pt}

\begin{footnotesize}{25 nm}\end{footnotesize} &\begin{footnotesize}  1.5 \end{footnotesize}& \cite{mei2015first}\\\hline
\end{tabular}
\end{center}
\end{table}
\subsubsection{Resonant Tunneling Diodes (RTD)}

 A resonant-tunneling diode (RTD)  operates according  to the tunneling principle, in which electrons pass through some resonant states at certain energy levels. RTD has been first demonstrated in 1974, where it consists of vertical stacking of nanometric epitaxial layers of semiconductor alloys forming a double barrier quantum well \cite{esaki1974new}, which allows the RTD to exhibit a wideband negative differential conductance \cite{alharbi2016high}. Over the last 10 years, progress has been achieved in increasing the output power of RTDs by almost two orders of magnitude and in extending the operation frequencies from earlier 0.7 THz to  values near 2 THz \cite{880842}.  Oscillations of RTDs in the microwave range were demonstrated at  low temperature in 1984\cite{sollner1984quantum} and the frequency was updated many times to several hundred GHz \cite{beer1994high}.  In 2010, a fundamental oscillation above 1 THz \cite{suzuki2010fundamental} have been attained.  The oscillation frequency was further increased up to 1.42 THz using thin barriers and quantum wells \cite{kanaya2014fundamental}. Further, the authors in \cite{maekawa2014frequency} and \cite{maekawa2016oscillation} indicated that reducing the length of the antenna integrated with the RTD  extended the frequency up to 1.55 THz and 1.92 THz, respectively.

RTD oscillators are actually suitable for wireless data transmission because the output power is easily modulated by the bias voltage and oscillations can be controlled by either electrical or optical signals. Wireless data transmission with a data rate of 34 Gbps has  been achieved in \cite{oshima2016wireless}. Because the size of RTD oscillators is small, it is possible to integrate multiple oscillators into one chip, which is convenient for multi-channel transmissions Indeed, wireless transmissions using both frequency division multiplexing (FDM) and polarization division multiplexing (PDM) have been demonstrated in \cite{oshima2017terahertz}, in which data rates up to 56 Gbps were obtained. Yet, the  drawback of this technology is that it cannot supply enough current for high power oscillations.  

\begin{figure}[h!] 
  \centering
  \includegraphics[width=3.6in]{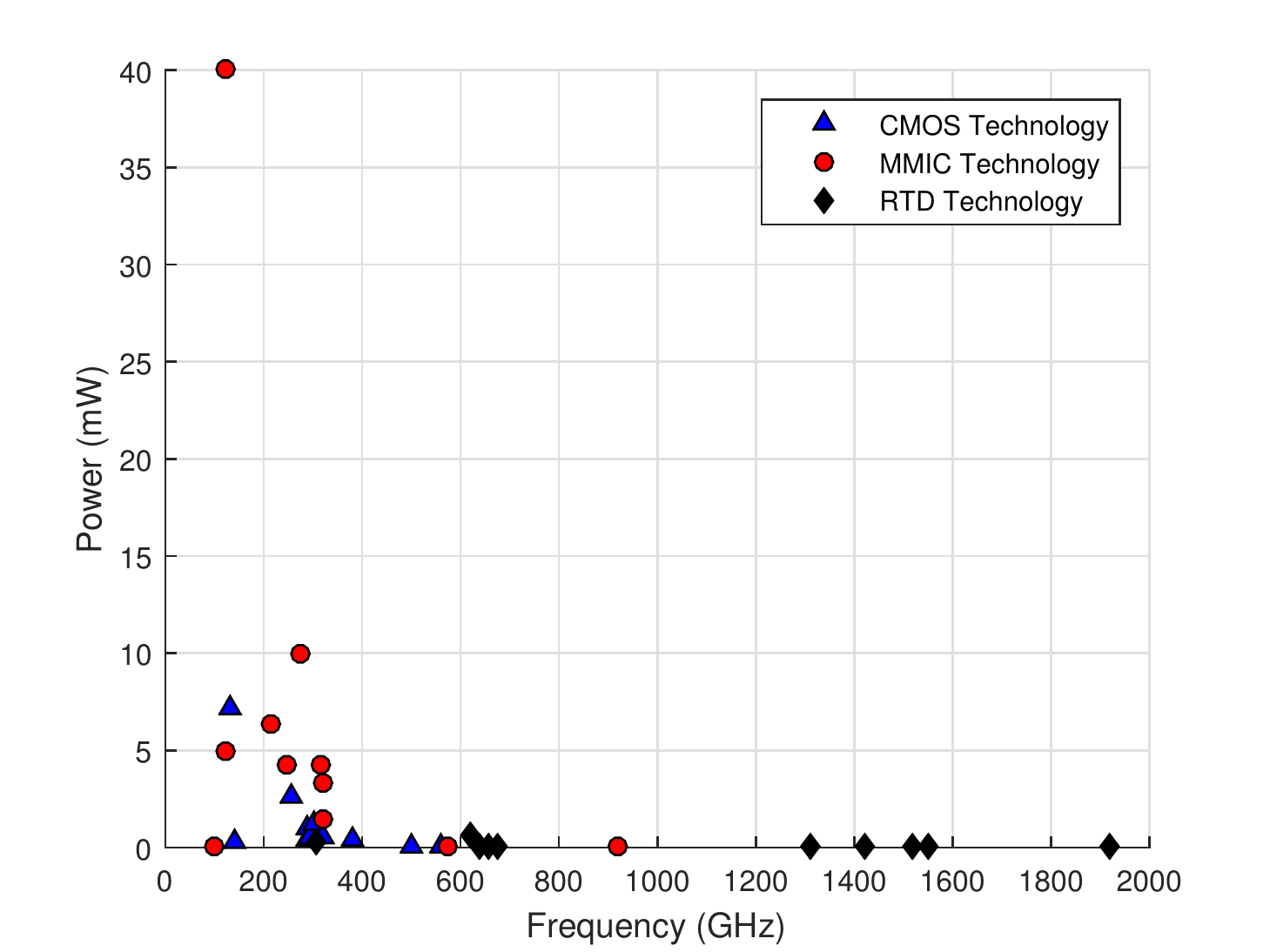}
  \caption{Solid-state electronics frequency of operation versus power.}
  \label{fig:All_tech}
\end{figure}

The technological progress that has been witnessed by the THz electronic devices is illustrated in Fig.~\ref{fig:All_tech}, where the frequency of operation for CMOS, MMIC and RTD technologies is displayed versus power. It could also be concluded that in the cases where  continued scaling of CMOS or integration with other silicon-based devices is inefficient, heterogeneous  as well as tunneling devices are deployed.  Nonetheless, despite the various progress that has been witnessed and is still ongoing in the field of solid state electronics,  the drastic power decrement associated with this technology is a major bottleneck. Thereby, other technologies have been gaining considerable attention.

\begin{table*}[!]
        \caption{Progress in THz Technology, achievable data rates and propagation distance}
        \begin{center}

                \label{table:Advances}
                \begin{tabular}{  c c c c c c c c c  }
                        \hline \hline \\ 
                        \textbf{Freq.} &   \textbf{ Data Rate} & \textbf{Spectral Eff.} & \textbf{Distance} & \textbf{Modulation} & \textbf{Power} & \textbf{Technology}   &  \textbf{Reference} & \textbf{Year}   \\  
                        (GHz)          & (Gbps)              & (bps/Hz)          &  (m) &  \textbf{Scehme} & (dBm) &   & & \\  \hline   
                        
        \\ 100  & 200 & - &  0.5 & QPSK &  -10  &Photodiode (PD)/ &  \cite{li2013400g}  &  2013 \\   &    &      &       &             &      &      subharmonic mixer (SHM) &                  &  \\

                        \\ 120  & 10 & 0.59 &  5800 & ASK and FEC &  16  &InGaAs/InP composite  &  \cite{hirata20105}  &  2010  \\ 
                        &    &      &       &             &      & channel HEMT MMIC    &                      &        \\

 \\ 120  & 10 &  -  &  5800 & ASK  &  16  & UTC-PD /InP HEMT MMIC  &  \cite{hirata2012120}  &  2012  \\ 
                        &    &      &       &             &      &     &                      &  \\
                     \\ 120  & 20 &  0.6  & 1700  & QPSK  &  7  &  InP  HEMT MMIC   &  \cite{takahashi2012120}  &  2012  \\ 
                        &    &      &       &             &      &     &                      &  \\

                        \\ 120  &   42/60 & 4.92/--      &   0.4  &  64QAM  & -10 &  GaAs HEMT  &          \cite{ando2016wireless}  & 2016 \\
                        
     \\ 130  & 11 &  -  & 3  &   ASK  &  8.6  &   CMOS Transceiver Chipset  &  \cite{fujishima2015tehrahertz}  &  2015  \\

  \\ 140  & 2/10                 & 2.86 & 1500 & 16 QAM & -5 & CMOS SHM/      &   \cite{wang201310} & 2013 \\
                        &   real/non-real time &      &     &        &    &                         Schottky barrier diodes  &                  &       \\

                        \\ 144 &  48  &  -   & 1.8 &  QPSK/QAM &  4 & Direct conversion I/Q/ & \cite{carpenter2016d}  & 2016\\
 &    &      &     &        &    &                        InP double HBT   &\\
                        
                        \\ 190 & 40/50 &  1/0.8  & 0.02/ 0.006 & BPSK &  -6  & 130 nm SiGe HBT & \cite{carpenter2016d} & 2017 \\ 
                          \\ 200  & 75 &  -  &  0.002 & QPSK  &   0    & UTC-PD &  \cite{shams2014photonic} & 2014 \\
 \\ 210  & 20 & 0.24   &  0.035 & OOK  &   4.6    & CMOS  &  \cite{wang2014cmos} & 2014 \\
                        \\ 220  & 25 &   0.74 &  0.5 & ASK  &  -3.4--1.4     & active MMIC (50-nm mHEMT) &  \cite{5971794} & 2011 \\
 \\  240 &  30   &   -  & 40  &   QPSK/8PSK  & -3.6       &  active MMIC components    &  \cite{6697568}   & 2013 \\

                        \\  240 &  64/96   &  1/1.5   & 40  &   QPSK/PSK  &   -3.6    &    MMIC  &  \cite{6956202}   & 2014 \\
  \\  240 &  64   &  -   & 850  &   QPSK/8PSK  &    -3.6   &    MMIC  &  \cite{kallfass201564}   & 2015 \\
                        
              \\  300 & 24 &    0.24  &   0.5  &  ASK &-7& UTC-PD/      & \cite{6248357}  &   2012  \\
                        &     &             &          &          &    &    Schottky  diode  &  &  \\ 
    \\  300 & 40 &   -   &  10   &  QPSK &-& Optical sub-harmonic IQ mixer       & \cite{kanno2015coherent}  &   2015  \\
                                           
                        \\  300  &  64  &  1  &  1   &   QPSK  &  -4   &  MMIC  &  \cite{kallfass2015towards} & 2015 \\

                        \\ 330  &  50 &  -  & 0.5$\sim$1   &  ASK                  &    -       & UTC-PD/  &   \cite{nagatsuma2015recent} & 2015\\
                        &     &             &          &          &    &    Schottky  diode detector &  &   \\           
                        
                        \\ 340  &  3 &   2.86 &  50  & 16 QAM through    32                   &   -17.5        & CMOS  SHM/  &   \cite{6693748} & 2014   \\
                        &    &        &      &  I/Q parallel channels             &                &  Schottky barrier diodes          &  \\
         
                        \\ 350-475  &  120 &  -  & 0.5  & QPSK                   &    -       & UTC-PD/  &   \cite{jia2017120} & 2017   \\
                        &    &        &      &               &                &  Schottky mixer         &  \\
         
           \\ 385  & 32  &   - &  0.5  & QPSK & -11           & UTC-PD/SHM  &   \cite{ducournau201532} & 2015    \\
           \\ 400  &  46 &   - &  2  & ASK                   &   -16.5        & UTC-PD/SHM  &   \cite{ducournau2014ultrawide} & 2014   \\
           \\ 400  &  60 &   - &  0.5  & QPSK                   &   -17        & UTC-PD/SHM  &   \cite{yu201560} & 2015   \\
      \\ 434  &  10 &  -  &  -  & ASK                   &   -18.5        & SiGe BiCMOS  &   \cite{hu2012sige} & 2012   \\
\\ 450  &  13 &  -  &  3.8  & QPSK   &   -28        &  photomixer/PD &   \cite{wang2018fiber} & 2018   \\
\\ 450  &  18 &  -  &   3.8 & PDM-QPSK   &      -28     &  photomixer/PD &   \cite{wang2018seamless} & 2018    \\
                
\\ 450  &  132 &  4.5  &  1.8  & QAM   &  -28         &  UTC-PD/photomixer &   \cite{li2019132} & 2019   \\
 \\ 542  &  3 &   - &  0.001  &     ASK   &     -6.7      & RTD  &   \cite{6198533} & 2012   \\
                      
           \\ 625  &  2.5 &   - &  3  &     ASK   &    -14       & Multiplier/SBD  &   \cite{moeller20112} & 2011   \\
                        &    &        &      &              &                &           &  \\
                        \hline 
                \end{tabular}
        \end{center}
\end{table*}

\subsection{Photonics Technologies}
THz devices based on electronic components possess both high resolution and high flexibility.  Yet, for many applications, THz measurements for wideband and high speed signals are needed. Such requirement may not be implementable via electronic devices due to the limited speed and bandwidth. However, modern photonics, which have been widely used for wideband and high speed microwave measurements can provide broader bandwidths \cite{yao2009microwave,minasian2013microwave}. 
In fact, the rise of THz wireless communication  began as early as the year 2000 upon the initiation of a 120 GHz wireless link  generated  by photonic technologies \cite{nagatsuma2000120}. The 120  GHz signal  was  the first commercial THz communication system  with an allocated bandwidth of 18 GHz. A data rate of 10 Gbps has been attained with an on-off keying (OOK) modulation and 20 Gbps  with a quadrature phase shift keying (QPSK) modulation \cite{hirata2012120,takahashi2012120}. This achievement attracted broadcasters who aimed to transmit high-definition TV data \cite{kleine2011review}  and demonstrated how photonic technologies played a key role in the development of first-age THz communication systems. Such achievement actually triggered the development of electronic devices and integrated circuits to strengthen the wireless technology. This eventually resulted in all electronic MMIC-based systems being successfully deployed in real-world events around the year 2008 \cite{hirata2012120}. Compared to solid-state electronics, photonic technologies   not only improves the data rate but also fuses both fiber-optics and wireless networks. These devices have broadband characteristics, high modulation index as well as high-speed amplitude and/or phase coding introduced from optical coherent network technologies \cite{nagatsuma2016advances}. The most fundamental and widely used devices are based on the optical-to-THz or THz-to-optical conversion using interaction media such as nonlinear optical materials, photoconductors, and photodiodes. High  speed  THz  wireless  communication  systems in  the  frequency range of 300 GHz-500 GHz, at data rates of 60 Gbps, 160 Gbps and up to 260 Gbps have been demonstrated in the literature indicating the potential of this  technology \cite{yu201560,yu2016160,pang2016260}. 

\subsubsection{Unitravelling Carrier Photodiode~(UTC-PD)}

The evolution of photonics technology greatly increased the speed of signal processing systems. Photodiodes  are examples of such devices that can provide both high speed and high saturation output resulting in the development of large-capacity communication systems. The combination of a high saturation power photodiode with an optical amplifier eliminates the post-amplification electronics, extends the bandwidth, and simplifies the receiver configuration \cite{653209}. In particular,  unitravelling carrier photodiodes  (UTC-PD) \cite{ishibashi1997uni} have a  unique mode of operation which makes them promising  candidates for such requirements. These photodiodes have been  reported  to  have  a  150 GHz  bandwidth \cite{sano1998ultra} and a high-saturation output current due to the reduced space charge effect in the depletion layer, which results from the high electron velocity  \cite{6861939}. Since the time UTC-PDs have been invented in 1997 \cite{ishibashi1997uni}, they have been used as photomixer chips. The frequency of the photomixer operation ranged from 75 to 170 GHz. Afterwards, the monolithic integration of a UTC-PD with  planar antennas was reported and the operation frequency exceeded 1 THz in 2003 \cite{ito2003photonic}. Upon antenna integration in UTC photomixers,  operation frequencies  exceeded 2 THz \cite{6861939}. UTC-PDs also enable the use of travelling-wave designs \cite{giboney1997traveling}, which provide  slower frequency response roll-off, and are more compatible with integration. UTC-PDs with output powers   of 148 $\mu$W at 457 GHz and 24 $\mu$W at 914 GHz have been approached \cite{renaud2006high}. In addition, a 160 Gbps THz wireless link has been achieved in the 300-500 GHz band using a single UTC-PD based transmitter as  shown in \cite{yu2016160}.
 
\subsubsection{Quantum Cascade Lasers (QCLs)}

A revolutionary advancement in THz technology arose in 2002 when successful operation of a quantum cascade laser (QCL) at THz frequencies has been reported in\cite{kohler2002terahertz}.  QCL basically bypasses semiconductor band-gap limitations in photonic devices by using sophisticated semiconductor heterostructure engineering and fabrication methods.  The semiconductor layers are thin, thereby,  very low energy transition happens when electron tunnel from  one layer to the other. Due to the low energy, the emitted radiation occurs in the THz region.  Since 2002, QCLs have quickly progressed  in frequency coverage, increased power output, and increased operating temperature. Currently, they are the only  sources capable of generating over 10 mW of coherent average power above 1 THz \cite{williams2006high}. In order to characterize the high modulation speed capability of THz QCLs and build a high speed THz communication link, a fast detector is necessary. The authors in \cite{5227008} demonstrated an all-photonic THz communication link at 3.8 THz by deploying QCL operating in pulse mode at the transmitter and a quantum well photodetector at the receiver. Later, the authors in \cite{6004763} were capable of increasing the frequency to 4.1 THz by using a QCL which operates in continuous wave mode.  

The progress witnessed in the photonics domain is a key enabler to the deployment of THz wireless links. Yet, the challenge remains in integrating these micrometer-scale bulky components of photonics into electronic chips. Surface plasmon-based circuits, which merge electronics and photonics at the nanoscale, may offer a solution to this size-compatibility problem \cite{leuthold2013plasmonic}. In  plasmonics, waves do not rely  on electrons or photons, but rather   electromagnetic waves  excite electrons  at  a  surface  of  a  metal and  oscillate at  optical  frequencies. An  advantage of these so-called surface plasmon polaritons (SPPs) is that they can be  confined to an ultra-compact area much  smaller  than  an  optical  wavelength. In addition, SPPs oscillate at optical frequencies  and  thus  can carry  information  at  optical  bandwidths.  The efficient wave localization up to mid-infrared frequencies  led plasmonics to become a promising alternative in future applications where both speed  and size matters \cite{7193578}. In particular, due to the two dimensional nature of the collective excitations, SPPs excited in graphene are confined much more strongly than those in conventional noble metals. The most important advantage of graphene would be the tunability of SPPs since the carrier densities in graphene can be easily controlled by electrical gating and doping. Therefore, graphene can be applied as THz metamaterial and can be tuned conveniently even for an encapsulated device \cite{luo2013plasmons}. Graphene-based THz components have actually shown very promising results in terms of generation, modulation as well as detection of THz waves \cite{hasan2016graphene}\cite{jornet2013graphene}\cite{7803807}. Furthermore, various unique generation techniques have been  recently proposed for THz waves. For instance, the authors in \cite{jin2017observation} experimentally demonstrated the generation of broadband THz waves from liquid water excited by femtosecond laser pulses. Their measurements showcased the significant dependence of the THz field on the relative position between the water film and the focal point of the laser beam. Compared with THz radiation generated from the air plasma, the THz radiation from liquid water has a distinct response to various optical pulse durations and shows linear energy dependence upon incident laser pulses. Such work will contribute to the exploration of laser-liquid interactions and their future as THz sources. Another example of  original THz generation techniques involves the work demonstrated in \cite{atakaramians2018enhanced}. The authors have shown that a dipole emitter can  excite the resonances of a nanofiber and lead to strong electric and/or magnetic responses.  They have experimentally demonstrated the magnetic dipole radiation enhancement for a structure containing a hole in a metallic screen and a dielectric subwavelength fiber. Their results are considered the first proof of concept of radiation enhancement of a magnetic dipole source in the vicinity of a subwavelength fiber. All these techniques will eventually result in breakthrough advancements in the various technological realms. 

As indicated by Table~\ref{table:Advances}, a tradeoff between power, distance and data rate have to be achieved in order to choose the most applicable THz wireless communication scenario based on the user requirements. By varying the modulation schemes from the most simple amplitude shift keying (ASK) to PDM-QPSK as well as experimenting with different configurations of THz-fiber integration, all electronics or all photonics systems, new opportunities are continuously developing for feasible  THz wireless communication scenarios.  

\begin{figure*}[htb]
\centering
\begin{tikzpicture}[
  level 1/.style={sibling distance=50mm},
  edge from parent/.style={->,draw},
  >=latex]
  
\node[root] {Terahertz Propagation Phenomenon}
  child {node[level 2] (c1) {Losses \cite{jornet2011channel}}}
  child {node[level 2] (c2) {Small-Scale Mobility  \cite{petrov2018effect} }}
  child {node[level 2] (c3) {NLoS Propagation \cite{han2015multi}}};

\begin{scope}[every node/.style={level 3}]
\node [below of = c1, xshift=23pt] (c11) {Spreading Loss};
\node [below of = c11, yshift=-2pt] (c12) {Absorption Loss };

\node [below of = c2, xshift=23pt] (c21) {micro-doppler effects};

\node [below of = c3, xshift=23pt] (c31) {Refraction};
\node [below of = c31] (c32) {Scattering};
\node [below of = c32] (c33) {Reflection};
\node [below of = c33] (c34) {Multipath};

\end{scope}

\begin{scope}[every node/.style={level 4}]
\node [below of = c12, xshift=20pt, , yshift=-2pt] (c121) {Water Molecules};

\node [below of = c34, xshift=15pt] (c341) {Small-Scale Fading};
\node [below of = c341, yshift=-5pt] (c342) {Large-Scale Fading};

\end{scope}

\foreach \value in {1,2}
  \draw[->] (c1.195) |- (c1\value.west);

\foreach \value in {1}
  \draw[->] (c2.195) |- (c2\value.west);

\foreach \value in {1,2,3,4}
  \draw[->] (c3.195) |- (c3\value.west);
  
\foreach \value in {1}
  \draw[->] (c12.195) |- (c12\value.west);

\foreach \value in {1,2}
  \draw[->] (c34.195) |- (c34\value.west);
    
\end{tikzpicture}

\caption{Terahertz Band Propagation Characteristics.}
\label{fig:channel_feat}
\end{figure*}
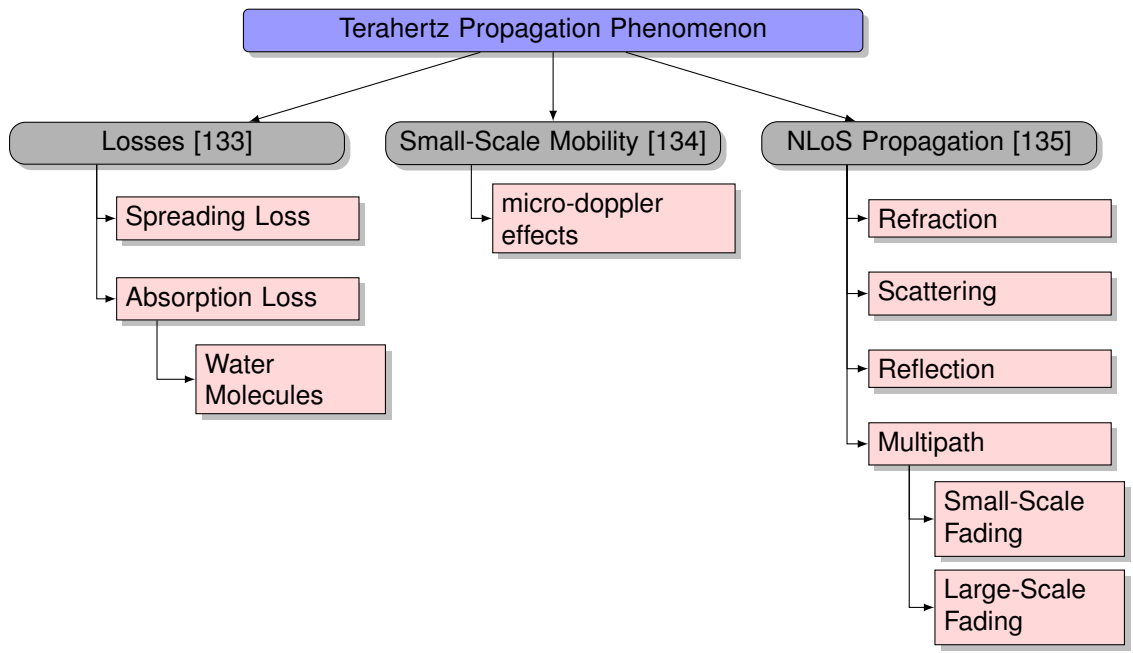

\section{Channel Modeling in the THz Frequency Band}
\label{Sec3}
\begin{figure*}[htb]
\centering
\begin{tikzpicture}[
  level 1/.style={sibling distance=50mm},
  edge from parent/.style={->,draw},
  >=latex]
  
\node[root] {Terahertz Channel Model}
  child {node[level 2] (c1) {Outdoor Channel Model  \cite{hirata2004high,hirata200910,hirata20105,hirata2012120,takahashi2012120}}}
  child {node[level 2] (c2) {Indoor Channel Model}}
  child {node[level 2] (c3) {Nanoscale Channel Model \cite{jornet2011channel,kokkoniemi2016discussion,kokkoniemi2015frequency,petrov2015interference,wang2017interference}}};

\begin{scope}[every node/.style={level 3}]
\node [below of = c1, xshift=15pt, yshift=-6pt] (c11) {Point-to-Point Links};

\node [below of = c2, xshift=26pt, yshift=-6pt] (c21) {Deterministic Models \cite{priebe2012calibrated,priebe2013ultra,moldovan2014and,han2015multi,sheikh2016terahertz,peng2016channel}};
\node [below of = c21, , yshift=-16pt] (c22) {Stochastic Models \cite{kim2016thz,priebe2013stochastic,kim2015statistical, kim2016statistical,saleh1987statistical,chong2003new,priebe2011aoa,choi2015performance,he2017stochastic,peng2015stochastic,8088634}};

\end{scope}

\foreach \value in {1}
  \draw[->] (c1.195) |- (c1\value.west);

\foreach \value in {1,2}
  \draw[->] (c2.195) |- (c2\value.west);
   
\end{tikzpicture}

\caption{Terahertz Channel Model Classification.}
\label{fig:channel_model}
\end{figure*}
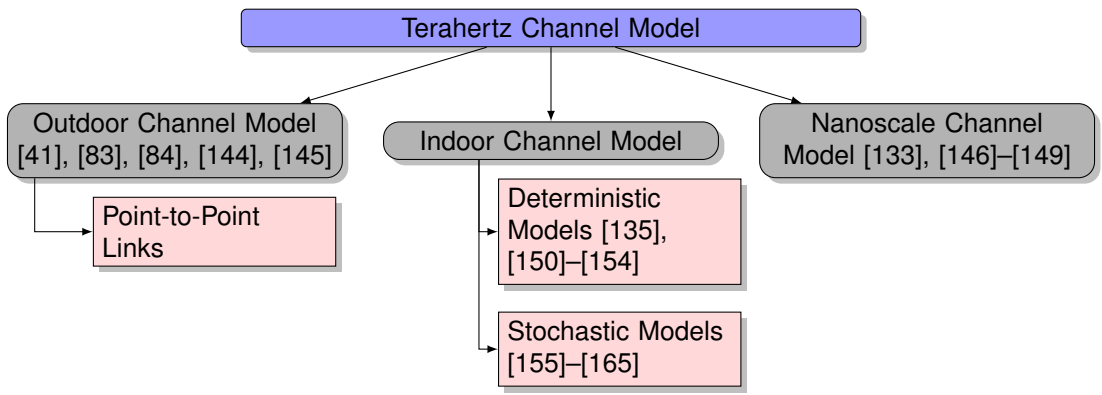

In order to realize an efficient wireless communication channel in the THz band, it is imperative to consider the various peculiarities  which distinguishes such frequency range. In fact, the THz frequency band has high frequency attenuation \cite{gordon2017hitran2016,jornet2011channel}, distinctive reflective \cite{nayar1991surface,jansen2008impact} and scattering \cite{piesiewicz2007scattering,4380579,jansen2011diffuse} properties as well as specular \cite{xu2002spatial} and non-specular \cite{priebe2011non} spatial distribution of the propagation paths. Moreover,  the highly directive antenna radiation pattern used to overcome high path loss results in frequent misalignments of beams due to small scale mobility  of user equipments \cite{petrov2018effect}.  The major propagation characteristics of THz waves are presented in  Fig. \ref{fig:channel_feat}. These effects cannot be neglected in the modeling process. As such, the existing channel models for the radio frequency (RF)  band cannot be reused for the THz band as they do not capture various effects including the attenuation and noise introduced by molecular absorption, the scattering from particles which are comparable in size to the very small wavelength of THz waves, or the scintillation of THz radiation. Such features  motivate the exploration of new models that efficiently characterize the  THz spectrum. In our discussion of channel modeling in the THz frequency band, we will follow the classification illustrated in Fig. \ref{fig:channel_model}.

\subsection{Outdoor Channel Models}

Models that emulate THz channels in outdoor environments are scarce focusing only on point to point links. The first 120 GHz experimental radio station license has been provided by the Ministry of Internal Affairs and Communications of Japan in 2004, where the first outdoor transmission experiments over a distance of 170 m have been conducted \cite{hirata2004high}. These experiments relied on utilizing mmW amplifiers along with high-gain antennas, such as  the Gaussian optic lens antennas or the Cassegrain antennas, leading to a successful outdoor transmission experiment. Starting from 2007 onward, the 120 GHz wireless signals were generated using InP HEMT MMIC technologies accounting on the electronic systems advantages of compactness and low cost \cite{hirata200910}. Upon the introduction of forward error correction (FEC)  technologies, a 5.8 km 10 Gbps data transmission was achieved by increasing both the output power as well as antenna gain \cite{hirata20105} \cite{hirata2012120}. The transmission data rate has been further increased to  22.2 Gbps by using the QPSK modulation scheme as shown in \cite{takahashi2012120}.

\begin{figure}[h!] 
  \centering
  \includegraphics[width=3.35in]{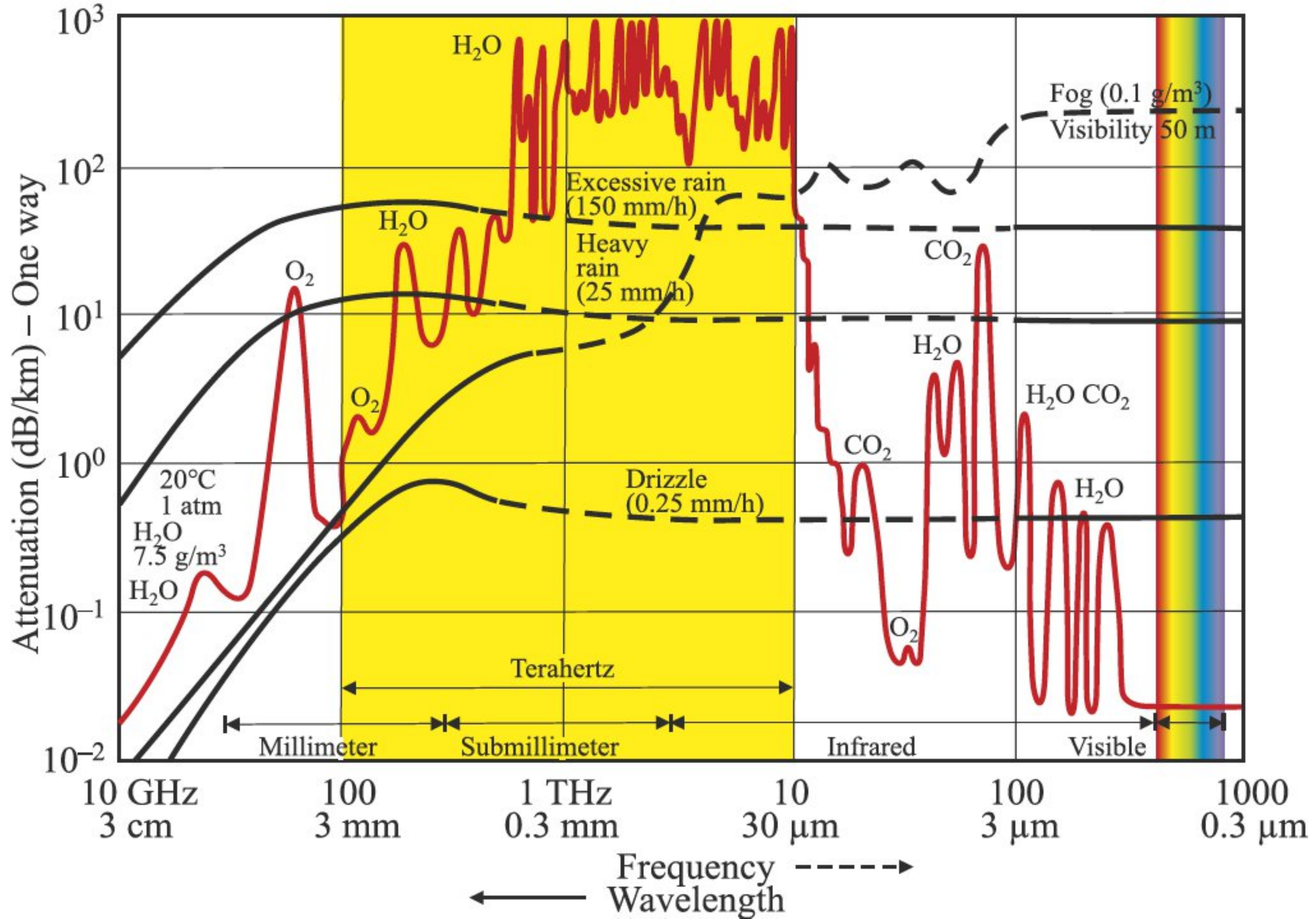}
  \caption{The attenuation impact of different environmental effects at various frequencies \cite{rogalski2011terahertz}.}
  \label{fig:Attenuation}
\end{figure}

The current outdoor channel models tackle only point to point cases. This is because few cases exist in the literature where experimental measurements have been reported. In specific, for outdoor measurements, the interference from unintentional NLoS paths can limit the bit error rate (BER) performance \cite{ma2018invited}. For long distance wireless communications, THz links can suffer significant signal loss due to atmospheric weather effects as illustrated in Fig. \ref{fig:Attenuation}. Yet, a closer look indicates that despite the
existence of absorption peaks centered at specific frequencies,
the availability of transmission windows allows establishing
viable communication at the THz frequency band. Thus, it will be important to estimate the weather impact on high capacity data links and compare the performance degradation of THz links in comparison to other competing wireless approaches \cite{federici2016review}.
As the  THz band channel is considered highly frequency selective, the transmission distance is limited by  attenuation and the appropriate carrier frequency  is determined according to the application. To be  capable of developing THz outdoor channel models, the evaluation of link performance using realistic data streams is needed. In our opinion,  a complete outdoor  channel model could be attained by  further exploring geometry-based, visibility-region based as well as map-based  models which include parameterization from measurement campaign  results. It must be emphasized that in order to operate  in outdoor environments, certain measures  have to be considered to avoid interference of passive services operating in the same band. Suitable frequency ranges are reported in Table \ref{table:fs} based on studies conducted in \cite{priebe2012interference,heile2015itu-r}.

\begin{table}[h!]
\caption{ THz Frequency Ranges for Fixed Services \cite{heile2015itu-r}}
\begin{center}
\centering
\label{table:fs}
\small
\begin{tabular}{|p{65pt}| p{65pt}|p{50pt}|}\hline

\raisebox{10pt}

\begin{footnotesize}\textbf{Frequency  Range (GHz)}\end{footnotesize}  & \begin{footnotesize}\textbf{Contiguous
Bandwidth (GHz) }\end{footnotesize}& \begin{footnotesize} \textbf{ Loss (dB/km)}\end{footnotesize}\\\hline

\raisebox{10pt}

\begin{footnotesize}275-320 \end{footnotesize}  &\begin{footnotesize} 
45\end{footnotesize}& $<$ 10\\\hline

\raisebox{10pt}

\begin{footnotesize}335-360 \end{footnotesize} &\begin{footnotesize} 25\end{footnotesize}& $<$ 10\\\hline

\raisebox{10pt}

\begin{footnotesize}275-370\end{footnotesize} & \begin{footnotesize} 95 \end{footnotesize}& $<$ 100\\\hline
 
 \raisebox{10pt}

\begin{footnotesize}380-445\end{footnotesize} & \begin{footnotesize} 65 \end{footnotesize}& $<$ 100\\\hline
 
  \raisebox{10pt}

\begin{footnotesize}455-525\end{footnotesize} & \begin{footnotesize}70\end{footnotesize}& $<$ 100\\\hline
 
  \raisebox{10pt}

\begin{footnotesize}625-725\end{footnotesize} & \begin{footnotesize} 100 \end{footnotesize}& $<$ 100\\\hline

 \raisebox{10pt}

\begin{footnotesize}780-910\end{footnotesize} & \begin{footnotesize}130 \end{footnotesize}& $<$ 100\\\hline
\end{tabular}
\end{center}
\end{table}

\subsection{Indoor Channel Models}

%

Unlike outdoor channel models, several indoor channel models are available in the literature. Indoor channel models can be categorized into either analytical or stochastic models. In terms of  deterministic channels, the  ray-tracing model is usually applied \cite{priebe2012calibrated,priebe2013ultra,moldovan2014and,han2015multi,sheikh2016terahertz,peng2016channel}. This  technique   is site-specific abiding with propagation theories and  capturing the phenomenon of wave transmission with precision as it is based  on geometrical optics, in which it is used for analyzing both the line of sight (LoS) and NLoS  THz wave propagation paths. Yet, the accuracy of the  ray-tracing models depends heavily on the complete knowledge of material properties. This   requires continuously adapting the model to a new environment, which can limit its  time efficiency. From the communications perspective, it is fundamental to understand the large  and small-scale statistics of the channel including path loss, shadowing and multipath propagation \cite{kim2016thz}. Hence, statistical methods arise as suitable options to  model THz propagation based on empirical channel measurements.  The first statistical model for THz channels, spanning the range between 275 and 325 GHz, has been provided in \cite{priebe2013stochastic}. The given model depends on  extensive ray-tracing simulations to realize the channel statistical parameters. Yet,  the information concerning the channel statistics such as the correlation function and power-delay profile cannot be captured easily. To tackle such concerns,  the authors in \cite{kim2015statistical, kim2016statistical} presented a geometrical statistical model for   device-to-device (D2D)  scatter channels at the sub-THz band. These models mimic the scattering and reflection patterns  in a sub-THz D2D environment. It is important to note that since the reflecting and scattering properties are frequency-dependent in the THz band, statistical distributions and parameters for intra-cluster and
inter-cluster need to be modeled properly. Therefore, a number of papers considered the characteristics of scattered multipath clusters including both angle and time of arrival for THz indoor channel modeling \cite{saleh1987statistical,chong2003new,priebe2011aoa}. In addition, by investigating the blocking probability in order  to describe the blocking effects of the propagation signals, the authors in \cite{choi2015performance} provided a modified THz channel model and proposed a path selection algorithm for finding the dominant signal. Similarly, in \cite{kokkoniemi2017stochastic}, the authors studied mean interference power and probability of outage in the THz band using stochastic geometry analysis. Further, the authors in \cite{tsujimura2017causal} presented a time-domain channel model in the THz band, where the coherence bandwidth has been computed for both the entire THz band and its sub-bands. The demonstrated numerical evaluation along with the provided experimental results indicate that the obtained impulse response satisfies causality and show that knowledge of the variations in the coherence bandwidth allows the selection of the proper center frequency for wireless communications in the THz band. Unlike traditional channel measurements, scenario-specific models are also available in the literature. The authors in \cite{he2017stochastic} presented a stochastic model for kiosk applications in the THz band, specifically between 220 and 340 GHz.  A 3D ray-tracing simulator has been utilized  to extract  channel characteristics of  three different kiosk application scenarios. Further, a stochastic channel model for future wireless THz data centers has been presented in \cite{peng2015stochastic}. The presented stochastic channel model accounts for the temporal and  spatial dispersion of the propagation paths and enables fast generation of channel realizations. Both the RMS delay and angular spreads are employed as a validation of the model. In \cite{8088634}, another  study on the statistical channel characterization of a THz scenario has been presented. This study  deals with the frequency range between  240 and 300 GHz and is considered one of the first to provide single-sweep THz measurement results. The measured data enables finer temporal details to be attained aiding the design of reliable transceiver systems including antenna misalignment problems. 

To achieve a balance between accuracy and efficiency, the authors in \cite{chen2018channel} suggested a hybrid channel model that combines both deterministic and statistical methods. In their discussion, the authors noted that a stochastic scatterer placement and ray-tracing hybrid approach could be developed. Scatterers in this case are stochastically placed, whereas the multipath propagation is traced and modeled based on ray-tracing techniques in a deterministic fashion. As such, geometry-based stochastic channel models are established.  The advantage of following such mechanism includes the high modeling accuracy and the low complexity. On the one hand, the very rich multipath effects are included using statistical modeling. On the other hand, the critical multipath components are computed deterministically.

Furthermore, the authors in \cite{zhao2019extending} provided an assessment for the  communication system design requirements at higher frequencies. In fact,  channel measurement results for 650 GHz carrier frequencies in comparison with 350 GHz carrier frequencies are given for a typical indoor environment. The  authors presented an extensive multipath channel model which describes the spatial distribution of all available paths with their respective power levels. Thereby, a more established perception is provided  for  THz wave propagation at different wavelength ranges.

\subsection{Nanoscale Channel Models}
In the past few years, advancements in the field of nanotechnology have paved the way towards the development of miniaturized sensing devices which  capitalize on the properties of novel nanomaterials. Such devices, denoted as nanodevices,  can perform simple tasks including computing, data storing, sensing and actuation. As  such, the formulation of nanonetworks will allow various applications in the biomedical, industrial, and military fields \cite{akyildiz2010electromagnetic}. Based on radiative transfer theory and in light of molecular absorption, a  physical channel model for wireless communication among nanodevices in the THz band is presented in \cite{jornet2011channel}. The provided model  considers  the contribution from the different types and concentrations of molecules, where the HITRAN database is used in order to compute the
attenuation that a wave suffers from. The Beer-Lambert law was used to compute the transmittance of the medium which relies on the medium absorption coefficient. The model provided in \cite{jornet2011channel} was also utilized to compute the channel capacity of nanonetworks operating in the THz band, in which the authors deployed different power allocation schemes. The authors recommended using the lower end of the THz band which has  lower absorption coefficients in order to ensure a strong received signal. Moreover, the sky noise model is  the basis of the existing absorption noise models. The authors in \cite{kokkoniemi2016discussion} elaborated on this topic by presenting different perspectives on how to model the molecular absorption noise. However, there is no  real experiments conducted in order to validate the proposed models. Not only absorption, but also scattering of molecules and small particles affects the propagation of electromagnetic waves. Hence, a wideband multiple scattering channel model for THz frequencies  has been demonstrated in \cite{kokkoniemi2015frequency}.   Further, the authors in \cite{petrov2015interference} presented an analytical model based on stochastic geometry for interference from omnidirectional nanosensors. However, in their model, they disregarded interference arising due to the existence of base stations. The authors in \cite{wang2017interference} tackled this issue where they studied interference from beamforming base stations. As such, it has been concluded that having a high density of base stations using beamforming  with small beam-width antennas and deploying a low density of nanosensors is recommended to improve the coverage probability.




\begin{table*}[!]
\caption{ Wireless Communication Candidates}
\begin{center}
\centering
\label{table:compare}
\small
\begin{tabular}{|p{70pt}| p{65pt}|p{65pt}|p{95pt}|p{90pt}|p{88pt}|}\hline

\raisebox{10pt}

\begin{footnotesize}\textbf{Technology}\end{footnotesize}&\begin{footnotesize} Millimeter Wave\end{footnotesize} & \begin{footnotesize}Terahertz\end{footnotesize}& \begin{footnotesize}Infrared\end{footnotesize} & \begin{footnotesize}Visible Light \end{footnotesize}&\begin{footnotesize}Ultra-Violet\end{footnotesize}\\\hline

\raisebox{10pt}

\begin{footnotesize}\textbf{Data Rate}\end{footnotesize}&\begin{footnotesize} Up to 10 Gbps\end{footnotesize} & \begin{footnotesize}Up to 100 Gbps\end{footnotesize} & \begin{footnotesize}Up to 10 Gbps\end{footnotesize}& \begin{footnotesize}Up to 10 Gbps\end{footnotesize} & \begin{footnotesize}Few Gbps\end{footnotesize}\\\hline
\raisebox{10pt}

\begin{footnotesize}\textbf{Range}\end{footnotesize} &\begin{footnotesize} Short range \end{footnotesize}  &\begin{footnotesize} Short range $-$ Medium range  \end{footnotesize}&\begin{footnotesize}Short range \
$-$ long range \end{footnotesize}& \begin{footnotesize}Short range\end{footnotesize}& \begin{footnotesize}Short range\end{footnotesize}\\\hline
\raisebox{10pt}

\begin{footnotesize}\textbf{Power Consumption} \end{footnotesize} &\begin{footnotesize} Medium\end{footnotesize} &\begin{footnotesize} Medium \end{footnotesize}& \begin{footnotesize}Relatively low\end{footnotesize}&\begin{footnotesize} Relatively low\end{footnotesize}& \begin{footnotesize}Expected to be low\end{footnotesize}\\\hline
\raisebox{10pt}

\begin{footnotesize}\textbf{Network Topology}\end{footnotesize} &\begin{footnotesize} Point to Multi-point\end{footnotesize} &\begin{footnotesize} Point to Multi-point  \end{footnotesize}& \begin{footnotesize}Point to Point\end{footnotesize}&\begin{footnotesize} Point to Point\end{footnotesize}& \begin{footnotesize}Point to Multi-point\end{footnotesize}\\\hline

\raisebox{10pt}

\begin{footnotesize}\textbf{Noise Source}\end{footnotesize} &\begin{footnotesize} Thermal noise\end{footnotesize} &\begin{footnotesize} Thermal noise  \end{footnotesize}& \begin{footnotesize}Sun Light + Ambient Light\end{footnotesize}&\begin{footnotesize}Sun Light + Ambient Light \end{footnotesize}& \begin{footnotesize}Sun Light + Ambient Light \end{footnotesize}\\\hline

\raisebox{10pt}

\begin{footnotesize}\textbf{Weather Conditions}\end{footnotesize} &\begin{footnotesize} Robust\end{footnotesize} &\begin{footnotesize}  Robust \end{footnotesize}& \begin{footnotesize}Sensitive\end{footnotesize}&\begin{footnotesize} $-$\end{footnotesize}& \begin{footnotesize}Sensitive\end{footnotesize}\\\hline

\raisebox{10pt}

\begin{footnotesize}\textbf{Security}\end{footnotesize} &\begin{footnotesize} Medium\end{footnotesize} &\begin{footnotesize}  High \end{footnotesize}& \begin{footnotesize}High\end{footnotesize}&\begin{footnotesize} High\end{footnotesize}& \begin{footnotesize}To be determined\end{footnotesize}\\\hline
\end{tabular}
\end{center}
\end{table*}

\section{Will the  Terahertz Band surpass Its Rivals ?}
\label{Sec4}

Carrier frequencies utilized for wireless communications have been increasing over the past years in an attempt to satisfy bandwidth requirements. While some of the interest of the research  community is steered towards the mmW frequencies in an attempt to fulfill the demands of next generation wireless networks, another direction involves moving towards optical wireless communication to allow higher data rates, improve physical security and avoid electromagnetic interference. The optical wireless connectivity is permitted using infrared, visible, or ultraviolet sub-bands, offering a wide range performance of coverage and data rate \cite{khalighi2014survey}. To highlight the necessity of utilizing the THz frequency band and showcase its capability in comparison to other envisioned enablers of future wireless communication, we present through the following subsections a comprehensive study of the features of the different technologies as summarized in Table \ref{table:compare}. 

\subsection{Millimeter Wave versus Terahertz}

Millimetre-wave frequencies of 28, 60 as well 73 GHz can enable myriad applications to existing and emerging wireless networking deployments. Recent researches introduced mmW as a new frontier for wireless communication supporting multiple Gbps within a coverage of few meters. The mmW frequency range has been adopted by  the Federal Communications Commission as the operational frequency of 5G technology.  By designating more bandwidth, faster, higher-quality video, and multimedia content and services will continue to be delivered\cite{rangan2014millimeter}. 

Despite the growing interest that arouse in mmW systems, the allocated bandwidth in such systems ranges from 7-9 GHz. This will eventually limit  the total throughput of the channel to an insufficient level due to consumers' increasing demand. Moreover, to reach the envisioned data rate of 100 Gbps, transmission schemes must have a challenging spectral efficiency of 14 bps/Hz \cite{kurner2014towards}.  In addition, the capacity of the fronthaul/backhaul link needed to achieve few Gbps should be several times higher than the user data rate to guarantee reliable and timely data delivery from multiple users. Nonetheless, as the frequency increases up to the THz band, Tbps links could be attained with moderate, realistic spectral efficiencies of few bits per second per Hz.
Operating at  the THz frequency band also allows a higher link directionality in comparison to mmW at the same transmitter aperture since THz waves have less free-space diffraction due to its shorter wavelength compared to the mmW.  Therefore, using small antennas with good directivity in THz communications  reduces both the transmitted power and the signal interference between different antennas \cite{ma2016terahertz}. Another interesting feature is the lower  eavesdropping chances in the THz band compared with the mmW. This is due to the high directionality of THz beams, which entail that unauthorized user(s) must be on the same narrow beamwidth to intercept messages.

\subsection{Infrared versus Terahertz}
One of the attractive, well-developed alternatives of radio frequency spectrum for wireless communication is the utilization of infrared radiation. The infrared technology uses laser transmitters with a wavelength span of 750-1600 nm that offer  a cost-effective  link with high data rates that could reach 10 Gbps. As such, it can provide a potential solution for the backhaul bottleneck \cite{uysal2014optical}. The infrared transmissions also  do not penetrate through walls or other opaque barriers, where they are confined to the room in which they originate. Such a feature secures the signal transmission against eavesdropping and precludes interference between links operating in different rooms. Nevertheless, as infrared radiation cannot penetrate walls, the installation of infrared access points that are interconnected via a wired backbone is required \cite{kahn1997wireless}. 

As  part of the optical spectrum, infrared communication faces similar  challenges that degrade its performance in different environments. For indoor environments, the ambient light signal sources, such as fluorescent lighting, induces noises at the receiver side. As for outdoor environments, in addition to moon/sun light noise level, atmospheric turbulence can limit the communication link availability and reliability, thus it is one of the main clogging factors of infrared communication deployment. The performance of optical links can be degraded even in clear weather as a result of scintillation,  and temporary spatial variation of light intensity.  Another major problem   is the necessity of developing pointing, acquisition, and tracking (PAT) techniques, which are essential for operation due to the unguided narrow beam propagation through the free space. As a result, optical transceivers must be simultaneously pointed at each other for communication to take place, in which precise alignment should be maintained \cite{fernandes1994wireless}.

THz frequency band is a good candidate to replace the infrared communication under inconvenient weather conditions such as fog, dust and turbulence. Fig.  \ref{fig:Attenuation} indicates that the THz band suffers lower attenuation due to fog compared to the infrared band. Recent experimental results showed that the atmospheric turbulence has a severe effect on the infrared signal, while it does not almost affect the THz signal. Moreover, the attenuation under the presence of cloud dust degrades the infrared channel but exhibits almost no measurable impact on the THz signal. As for the noise, THz systems are not affected by ambient optical signal sources. Due to the low level of photon energies at THz frequencies, the  contribution to the total noise arises from the thermal one \cite{rogalski2011terahertz}.

\subsection{Visible Light versus Terahertz}
Communication through visible light is a promising energy-aware technology that has attracted people  from both industry and academy to investigate its potential applications in different fields.  Visible light communication (VLC) carries information by modulating light in the visible spectrum (390-750 nm) \cite{arnon2015visible}. Recent advancements in lighting through light emitting diodes (LEDs) have enabled unprecedented energy efficiency and luminaire life span since LEDs can be pulsed at very high speeds without noticeable effect on the lighting output and human eye. LEDs also possess several attractive features including their low power consumption, small size, long life, low cost, and low heat radiation. Therefore, VLC can support a lot of vital services and application such as indoor localization, human-computer interaction, device-to-device communication, vehicular networks, traffic lights, and advertisement displays~\cite{khalighi2014survey}.

Despite the advantages associated with the deployment of VLC communication, several challenges exist that could hamper the effectiveness of the wireless communication link. In order to achieve high data rates in VLC links, a LoS channel should be primarily assumed in which both the transmitter and the receiver ought to have aligned field of views (FOV) to maximize the channel gain. Nevertheless, due to the receiver movement and continuous changes in orientation, the receivers' FOV cannot be always aligned with the transmitter. Such misalignment results in a significant drop in the received optical power \cite{pathak2015visible}. In occasions where an object or a human blocks the LoS, a noticeable degradation of the optical power is witnessed resulting in severe data rate reduction. Similar to infrared waves, interference from ambient light can significantly reduce the received signal to oise ratio (SNR), degrading the communication quality \cite{arnon2015visible}.  Current research in visible light networking also sheds the light on downlink traffic without taking into consideration how the uplink can operate. Since a directional beam towards the receiver should be maintained in VLC uplink communication, significant throughput reductions when the mobile device is constantly moving/rotating may occur. Thus, other wireless technology should be used for transmitting uplink data \cite{khalighi2014survey}. 

Contrary to VLC systems, the THz frequency band permits NLoS propagation, which acts as a supplement when LoS is unavailable \cite{akyildiz2014terahertz}. In such scenarios, NLoS propagation can be designed by strategically placing mounted dielectric mirrors to reflect the beam to the receiver.
The resulting path loss is adequate due to the low reflection loss on dielectric mirrors. In fact, for distances up to 1 meter and a transmit power of 1 Watt, the capacity of only the NLoS component of a THz link is around 100 Gbps \cite{moldovan2014and}. Furthermore, the THz frequency band is considered a candidate for uplink communication, a capability which VLC communication lacks. Another specific application where THz becomes a valuable solution is when there is a need to switch the lights off while looking for network service. Due to the restriction of positive and real signals, VLC systems will suffer from spectral efficiency loss. Indeed, utilizing unipolar  OFDM system by imposing Hermitian symmetry characteristic  leads to 3 dB performance loss in comparison to traditional bipolar systems that can be used in THz communication \cite{wang2017optical}.

\subsection{ Ultra-Violet versus Terahertz}
To relax the restrictions enforced by the PAT requirements of optical wireless communication, researchers investigated the optical wireless communication with NLoS capabilities. The deep ultra-violet (UV) band (200-280 nm) proves to be a natural candidate for short range NLoS communication, which is known also as optical scattering communication. In fact, since solar radiation is negligible at the ground level,  the  effect of background noise  is  insignificant,  allowing  the  use  of  receivers  with  wide FOV. Thus, NLoS-UV can be used as an alternative to outdoor infrared or VLC links or in combination with existing optical and RF links  as it is relatively robust to meteorological
conditions \cite{uysal2014optical}.

Although UV communication possesses favorable features, it suffers from a number of shortcomings. For LoS links and despite the deployment of  moderate FOV receivers, achievable ranges are still limited due to absorption by the ambient ozone. When operating under NLoS conditions for long ranges, the detrimental effects of fully coupled scattering as well as turbulence deteriorate the communication link. The effect of fading further impacts the received signal resulting in a distorted wave-front  and fluctuating intensity. Therefore,  data rates are limited to few Gbps and distances are restricted to short ranges\cite{xu2008ultraviolet}.

Compared to UV links, the THz frequency band is considered a suitable contender.   Unlike UV communication which imposes health restrictions and safety limits on both the eye and skin, an important point to emphasize is that the THz band is a non-ionization band; therefore, no health risks are associated with such frequencies \cite{pickwell2006biomedical}. From a communication perspective, this indicates that the THz data rates will not be vulnerable to any constraints. The fact that developing a UV system model suitable for practical application scenarios is still a demanding issue indicates that THz can compete  UV communication in its anticipated applications.

\section{Terahertz Applications}
\label{Sec5}
\begin{figure*}[!] 
  \centering
  \includegraphics[width=4.7 in]{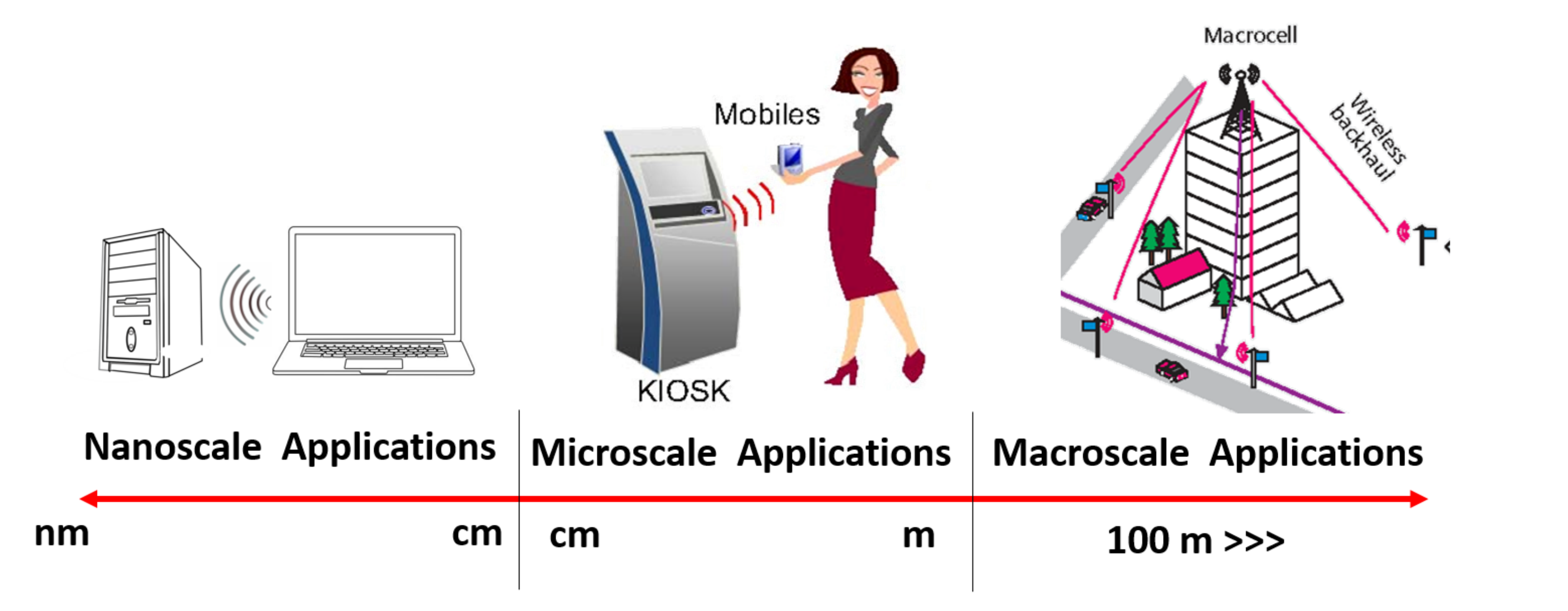}
  \caption{Different ranges of Terahertz related applications.}
  \label{fig:scales}
\end{figure*}

The THz band is envisioned as a potential candidate for a plethora of applications, which exist within the nano, micro as well as macro scales as illustrated in Fig. \ref{fig:scales}. Tbps data rates, reliable transmission and minimal latency \cite{boulogeorgos2018terahertz} are among the multiple features that allow such band to support  several scenarios in diverse domains.
\subsection{Terahertz Nanoscale   Applications }
On a nanoscale and with the advent of the Internet of NanoThings (IoNT), the interconnection of various objects, sensors as well as devices results in ubiquitous networks tailored not only for device-to-device communication but also for extracting data from areas hard to access. Based on such technological progress,  the communication architecture of nanonetworks has been established.  These networks rely on the THz band to achieve communication between its different entities constituting of nanoscale transistors, processors as well as memories \cite{akyildiz2010internet}. The interconnection of these pervasively deployed nanodevices with existing communication networks via the Internet creates a cyber physical system. Thus, nanoscale wireless communication is a key enabler of applications involving operations inside computers and devices for a typical range of few cm. These include chip-to-chip, board-to-board and device-to-device communications. In addition, THz nanocells are envisioned to be part of the hierarchical cellular network for potential mobile users to support various indoor as well as outdoor applications \cite{kurnerapp}. Actually, almost all modern automation depends on nanoscale devices that can communicate with each other in order to provide smarter technical options. Hence, nanoscale communication is suited for applications in multimedia, security and defense, environment and industry as  well as biomedical applications \cite{jornet2012internet}.
 For example, THz nanosenors, detectors and cameras can support security applications through the capabilities that  THz radiation possess which  enables the detection of weapons, explosives as well as chemical and biological agents \cite{federici2005thz}. From an environmental perspective, THz nanosensors allows the detection of pollutants and as such renders the technology useful for food preservation and food processing applications. In terms of imaging, the THz band spectroscopic characteristics  surpasses the currently available backscattering techniques and elucidates the dynamics of large biomolecules \cite{plusquellic2007terahertz}. In addition, nanoantennas enable wireless interconnection amongst nanosensors deployed inside and over the human body resulting in many bio-nanosensing applications \cite{yang2016channel}. Several works exist pointing to the THz band as an enabler of in-vivo wireless nanosensor networks (iWNSNs) \cite{nafari2015metallic}\cite{biagioni2012nanoantennas}.
In particular, the authors in \cite{elayan2017terahertz} presented an attenuation model of intrabody THz propagation to facilitate the accurate design and practical deployment of iWNSNs. In subsequent studies, the authors also demonstrated both the photothermal impact \cite{elayan2017photothermal} along with the noise effect \cite{elayan2018end} of THz intrabody communication to further verify the feasibility and prosperity of such propagation  mechanism.

\subsection{Terahertz Microscale  Applications}
 THz wireless communication promises luring applications that meet consumer's demands of higher data rates especially at the micro-scale.  Wireless local area network (WLAN) and wireless personal area network (WPAN) form the basis of such applications which include high-definition television (HDTV) in home distribution, wireless displays, seamless transfer of files, and THz access points in the areas with human congestion. The THz band provides small cell communication for mobile cellular networks, where ultra-high data rate can be provided to mobile users within transmission range up to 20 m.  As such, THz frequencies provides transmission solutions in adhoc networks and for nomadic users by facilitating connection  to access points including gates to the metro station, public building entrances, shopping malls, etc. In addition, microscale wireless communication at the THz band involves wireless transmission of uncompressed high definition (HD) videos  for education, entertainment, telemedicine, as well as security purposes. The authors in \cite{8422216} actually demonstrated the integration of a 4K camera into a THz communication link and showed the live streaming and recording of the uncompressed HD and 4K videos, followed by analysis of the link quality. The BER was measured at several link distances, where even at the maximum distance of 175  cm, the BER was below the FEC limit of $10^{-3}$. Not only that, NHK (Japan broadcasting corporation) has already started trial experiments by telecasting 8K video using proprietary devices for Olympic games that will be held in 2020 \cite{miki2015ready}. Within the same scope, the new vision of modern railways signifies the need to interconnect infrastructure, trains and travelers. Therefore, to realize a seamless high data rate wireless connectivity, huge bandwidth is required. Such demand motivates the deployment of  THz communications as they can offer orders of magnitude greater bandwidth than current spectrum allocations and enable very large antenna arrays which in turn provide high beamforming gains \cite{kurner2016millimeter}. This facilitates relevant  scenarios for railway applications including train to infrastructure, inter-wagon and intra-wagon communications. Further, kiosk downloading is another example of microscale application at THz  frequencies, which offers ultra high downloads of digital information to users' handheld devices. For instance, Ad posters in metros, trains or streets can be the front interface for downloading pre-fixed contents such as newly released movie trailers, CDs, books, and magazines\cite{kurnerapp}.

\subsection{Terahertz Macroscale Applications} 
 
On a macroscale, THz wireless communication facilitates potential outdoor applications which range from few meters up to kilometers. For instance, wireless backhauling/fronthauling is one of the envisioned applications for the standard  100 Gbps transmission solutions \cite{kurnerback}. In terms of backhauling, wireless  point-to-point links are  widely  applied  for  transmission  of  information  to the base stations of macrocells especially in those points where optic fiber is not available. In terms of fronthauling, wireless  point-to-point links are those between the radio equipment controller of a base station and the remote radio head (radio unit). These systems are normally  operating  within the spectrum of 6 GHz  to 80 GHz, in which they necessitate  strict  compliance  with  the LoS  conditions between  the  transceivers  of  two  nodes \cite{narytnyk2014possibilities}. The increasing number of mobile and fixed users in both the private, industrial and service sectors will require hundreds of Gbps in the communication either to or between cell towers (backhaul) or between cell towers and remote radio heads (fronthaul). In such scenarios, apart from the high targeted data-rates (1 Tbps), the critical parameter is range, which should be in the order of some kilometers \cite{boulogeorgos2018terahertz}. From the point of  view of economic feasibility, the principal  difference between the microwave solutions  and the solutions for THz waves covers the price of spectrum, equipment costs and the difference in the time spent for assembly and on-site tuning. Future advancements which include massive deployment of small cells,  implementation of cooperative multipoint transmission  and Cloud Radio Access Networks (C-RAN) may increase the required data rates for either fronthauling or backhauling or both.   
 
  Wireless data centers are considered another promising application at the macroscale. Actually, the increasing call  for cloud applications triggered competition between data centers in an attempt  to  supply users with  an upgraded experience. This is accomplished by accommodating an extensive number of servers and providing adequate bandwidths to support many applications. In fact, wireless networking possess several features including the the adaptability and efficiency needed to provide possible ways to manage traffic bursts and finite network interfaces \cite{6108333}. Nonetheless, wireless transmission capabilities are limited to short distances and intolerance to blockage leading to a deterioration in the efficiency of data centers if all wires are substituted. A better alternative exists through the  augmentation of the data center network with wireless flyways  rather than exchanging all cables  \cite{halperin2011augmenting}. The authors in \cite{7502886} suggested using  THz links in  data centers as a parallel technology. Such deployment in data centers results in an enhanced  performance experience  along with immense savings in cable prices  without compromising any throughput. The authors adopted  a bandwidth of 120 GHz for  data center applications, where  atmospheric data has been utilized to model the THz channel.
\section{THz Standardization Activity}
\label{Sec6}
The work towards developing a powerful THz standard has launched during the last decade when the THz communication research was still in its infancy stage. In 2008, the IEEE 802.15 established the  \textit{THz Interest Group}  as a milestone towards investigating the operation in the so called ``no man's land'' and specifically for frequency bands up to 3000 GHz. The new group conducted a liaison to the International Telecommunication Union (ITU) and the International Radio Amateur Union (IARU) regarding the description of the frequency bands higher than 275 GHz. Moreover, the group launched a call for contribution to cover different topics including possible THz applications, ways to realize transmitters and receivers, expected ranges and data rates, impact on regulations and market as well as ongoing research status. The journey of THz exploration started with studying the link budget for short distances considering the atmospheric attenuation for frequencies up to 2 THz. Despite the uncertainty in determining the realistic transmitted power, receiver sensitivity and thermal noise floor at this band, the study concluded the THz potential to deliver multi Gbps at an early time in 2008 \cite{roberts2008link}. Then, further solid analysis were conducted based on Shannon theory principles to prove the THz applicability for future in-home application with a data rate of 100 Gbps \cite{razoumov2008feasibility}. In addition, the THz interest group discussed the recent advances in research and lab measurements that encourage investigating the 300 GHz radio channel \cite{kurner2008terahertz}. Specifically, detailed discussions about the current status of semiconductor technologies and photons based techniques for generation have been conducted in \cite{rieh2008current, mearini2008high} and \cite{ridgway2008millimeter, liu2008integrated}, respectively. Another important aspect that has been discussed is the desirable performance to the industry in addition to the cost and safety issues  \cite{roberts2008some, roberts2008some2}.

In Nov. 2008, a science committee has been formed in order to bring the THz science communities together as a step to convert the THz interest group to a study group.  To that end, the committee provided a comprehensive study on channel models, gave a general overview of technology trends and provided helpful technical feedback to ITU \cite{kurner2009scope}. In March 2010, the THz interest group renewed the THz call for contributions to discuss the advances since the last call and further investigate the applicable modulation techniques, THz channel models, THz needed infrastructure and several other points \cite{kurner2010towards}. In Nov. 2010, the interest group discussed the issues that will enable the THz communication deployment in order to prepare the agenda of  the next ITU WRC that would be held in 2012 \cite{kurner2010towards}. The discussion included defining spectrum bands for active services, where several bandwidths are defined with allowable attenuation for short distances.  Moreover,   the discussion showed the necessity  to develop a holistic design approach which includes  investigating channel characteristics by measurements, designing antennas to overcome  the high attenuation, defining suitable communication systems, building  an integrated RF front end and consider the connection to backbone network. In 2011, the THz interest group put more effort on investigating the existing THz generation technologies and the potential communication performance in addition to the expected road map in order to be discussed in WRC 2012 \cite{kurner2011update, hosako2011the, roberts2011some}. 
In March 2012, the interest group reviewed the results of WRC 2012 and the ITU radio regulations which allow the coexistence of active services beside  passive services in the frequency band 275-1000 GHz. Specifically, the radio astronomy service occupies  275-323 GHz, 327-371 GHz, and 388-424 GHz, while the earth exploration-satellite and space research services operates in 
275-277 GHz, 294-306 GHz and 316-334 GHz bands. The main issue in the discussion was about the necessary practical steps that should be adopted to prevent various active services (nomadic links, fixed links, airborne systems and multiple interferes) from interfering with the aforementioned passive services \cite{kurner2012review, briebe2012will}. The interest group discussed the prerequisites needed to start a study group, which included  the participation of MAC expertise and people from industry in addition to the current PHY contributions \cite{kurner2012on}.

Staring from 2013, the interest group added the MAC layer to its discussion sessions in order to investigate the requirements that should be fulfilled by the MAC protocols to accommodate for several THz communication applications \cite{priebe2012mac}. A link level study is conducted via a simulation environment for THz communications using ray-tracing channel model \cite{rey2012link}.  Moreover, the data center operation and requirements have been discussed  as a guide for future THz utilization for data center interconnection links \cite{kurner2012literature, kasamatsu2012optical, yunlong2012thz}. Up until this stage of time, the IEEE 802.15 THz interest group activities included introducing a summary of THz technological developments,  channel modeling and spectrum issues as well as working to generate a technical expectations document \cite{kurner2013on}. In July 2013, the THz interest group proposed starting a study group to explore the possibility of launching a standard towards 100 Gbps over beam switchable wireless point-to-point links, which can be used in wireless data center and backchaining. The inauguration of IEEE 802.15 study group 100G has been done in September 2013 \cite{kurner2013inuaguration}. The study group working tasks included discussing current technologies limits, investigating relevant PHY and MAC protocols, defining possible applications  and introducing proposals for THz communication on wireless data centers \cite{kurner2013on}.

 In 2014, a group called   ``the task group 3d (TG3d)" has been initiated to adjust the 802.15.3 metrics in an aim to address100 Gbps  for switched point-to-point links. Several  applications are involved within this category   including wireless data centers,  backhauling/fronthauling as well as close-proximity communication such as kiosk downloading and D2D communication \cite{kurner2014tg3d}. The first step towards defining bands for active services has been done when IEEE contacted the ITU to discuss allocating the THz band from 275 GHz to 325 GHz for  mobile and fixed services. The ``spectrum engineering techniques'' ITU group confirmed also the availability of 23 GHz in the band 252-275 GHz for mobile and fixed services \cite{heile2015itu-r}. In addition, the WRC 2015 agreed to discuss the land-mobile and fixed active services spectrum allocation in 275-450 GHz while maintaining protection of the passive services in the agenda of WRC 2019 \cite{rey2016progress}.  To this end, the ITU-R is invited to identify technical and operational characteristics, study spectrum needs, develop propagation models, conduct sharing studies wit` the passive services and identify candidate frequency bands. Specifically, 8 groups namely: spectrum engineering techniques, propagation fundamentals, 
 point-to-area propagation, point-to-point and earth space propagation, land mobile service, fixed services, space research, earth exploration-satellite service, and radio astronomy, are involved in conducting these studies \cite{rey2016progress}. The first standard of THz communication came to the scene in 2017, where it focused on point-to-point  highly-directive links using 8 different channel bandwidths (as multiples of 2.16 GHz) \cite{kurner2018ieee}. Within the past two years, the interest group discussed several THz research activities such as multi-scale channel measurements, statistical channel characterization, solid state generation methods, antenna array designs, THz networks challenges and design, interference studies for THz intra-device communication systems and measurements of research data center.

\begin{table}
        \caption{Timeline of THz Standardization.}
        \centering
        \begin{minipage}[t]{.7\linewidth}
                \color{gray}
                \rule{\linewidth}{1pt}
                \ytl{\textcolor[rgb]{0,0.25098,0.501961}{Jan. 2008}}{the IEEE 802.15 established the ``THz Interest Group"}
                \ytl {\textcolor[rgb]{0,0.25098,0.501961}{Mar. 2008}}{Call for contribution}
                \ytl {\textcolor[rgb]{0,0.25098,0.501961}{Nov. 2008}}{Science committee formation}
                \ytl {\textcolor[rgb]{0,0.25098,0.501961}{Jul. 2009}}{Call for THz application}
                \ytl {\textcolor[rgb]{0,0.25098,0.501961}{Nov. 2011}}{Call for THz application}
        \ytl {\textcolor[rgb]{0,0.25098,0.501961}{2012}}{THz applications and PHY layer issues discussion}
        \ytl {\textcolor[rgb]{0,0.25098,0.501961}{Sep. 2013}}{Inauguration of IEEE 802.15 study group 100G}
        \ytl {\textcolor[rgb]{0,0.25098,0.501961}{Dec. 2013}}{Study group 100G call for applications}
        \ytl {\textcolor[rgb]{0,0.25098,0.501961}{2014}}{Task group 3d (TG3d) formation}
        \ytl {\textcolor[rgb]{0,0.25098,0.501961}{2015-2016}}{Discussion with ITU about THz band allocation for mobile and fixed services}
        \ytl {\textcolor[rgb]{0,0.25098,0.501961}{Sep. 2017}}{IEEE Std 802.15.3d-2017 standard is approved as 100 Gbps wireless switched point-to-point system}
                \bigskip
                \rule{\linewidth}{1pt}%
        \end{minipage}%
\end{table}

\section{Future Research Directions}
\label{Sec7}
In this section, we shed the light on key enablers that will facilitate the progress and deployment of THz frequency links  as well as open the door towards numerous applications that support both cellular as well as vehicular networks.  
\subsection{Terahertz Ultra-Massive MIMO} 
The THz frequency band is considered a key enabler in satisfying the continuously expanding demands of higher data rates. Yet, despite the huge bandwidth it provides, the band suffers from high atmospheric losses. Therefore, high-gain directional antennas are utilized in order to invoke communication over distances exceeding a few meters. Specifically, in the THz band, antennas become  smaller and  more  elements can be installed in the same footprint. As such, stemmed from the Massive MIMO concept \cite{larsson2014massive}, the authors in \cite{akyildiz2016realizing} formulated an Ultra-Massive MIMO (UM-MIMO) channel. The concept of  UM-MIMO  relies on the adoption of ultra-dense frequency-tunable plasmonic nano-antenna arrays which are simultaneously utilized in transmission and reception thereby increasing the communication distance and, ultimately, the achievable data rates at THz frequencies \cite{zakrajsek2017design}. Actually, the radiated signals may be regulated both in the elevation and the azimuth directions when securing two-dimensional or planar antenna arrays rather than one-dimensional or linear arrays. This results in   3D or Full-Dimension MIMO. The performance of  UM-MIMO technology depends on two metrics, namely, the prospects of the plasmonic nanoantenna as well as the characteristics of the THz channel. As such, a channel model for the UM-MIMO systems using the array-of-subarray architecture has been proposed in \cite{han2018ultra}. The results indicate that when using 1024 $\times$ 1024 UM-MIMO systems at 0.3 THz and 1 THz,  multi-Tbps links are achievable at distances of up to 20 m. Another important aspect is the dynamic resource allocation that can fully utilize the UM-MIMO system and gain the maximum benefits by adaptive design schemes \cite{rodrigo2018multi}. Further,  spatial modulation  techniques  that  can  influence  the  attributes of  densely  packed  configurable nanoantenna subarrays have been studied by the authors in \cite{sarieddeen2019terahertz}. By using such an approach, both the capacity  and  spectral  efficiency of the system are improved while  maintaining  acceptable beamforming  performance.   A  particular spatial modulation  configuration that establish good channel conditions is suggested based on  the  communication distance  and  the  frequency  of  operation.

\subsection{Terahertz Virtual  Reality Perception via Cellular Networks}
In order to attain a high-mobility  automotive  content  streaming guarantee and guarantee an ultra reliable, low latency communication, it is essential to  go  well  beyond  what  5G  can  deliver. Although  there  are  numerous  compelling  augmented reality and   virtual reality applications,  video  is  the  most  important  and  unique  in  its  high  bandwidth  requirements.  As such, the THz frequency band is sought  as a technology that  will provide both   high capacity   and dense coverage  to bring these   applications close to the end user.  THz cellular networks will  enable interactive, high dynamic  range videos at  increased  resolutions and  higher  framerates, which actually necessitate 10 times  the  bit-rate  required  for  4K   videos.  THz transmission will help relieve any interference problem and provide extra data to support various instructions  in video transmission. In addition, the THz band will be an enabler of 6 degrees of freedom (6DoF) videos   providing  users with an ability  to  move within and  interact with the environment. Streaming live 6DoF content to deliver a “be there” experience is basically a   forward-looking  use  case \cite{WinNT}. The results presented in \cite{chaccour2019reliability} show that THz can deliver rates up to 16.4 Gbps with a delay threshold of 30 ms given that the impact  of molecular absorption on the THz links, which considerably limits the communication range of the small base station, is relieved through network densification.
\subsection{Terahertz Communications for Mobile HetNets}

\begin{figure}[h!] 
  \centering
  \includegraphics[width=3.45in]{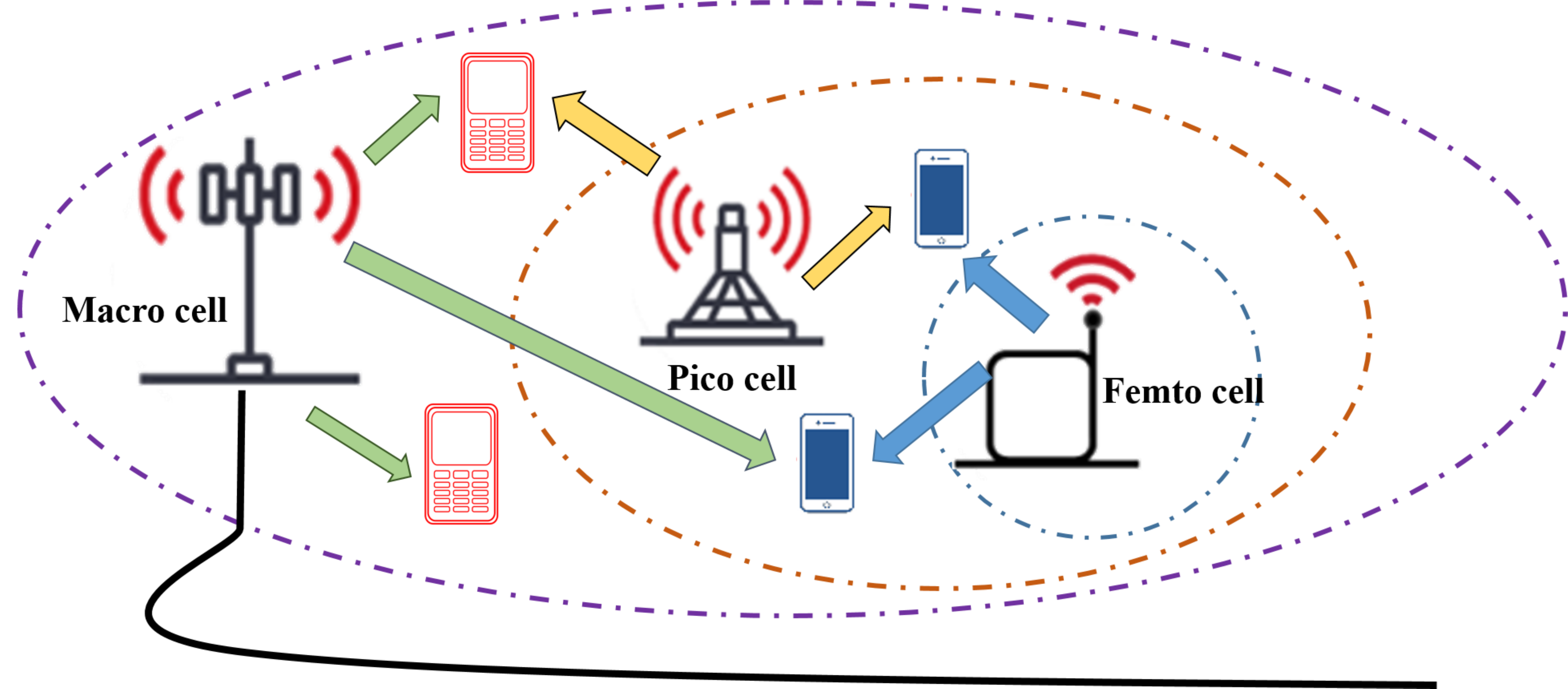}
  \caption{Picocells  and  femtocells will be  collocated  within  the  macrocell footprint for THz wireless communication.}
  \label{fig:hetnet}
\end{figure}

As the demands of communication services are developing in the direction of multiple users, large capacity and high speed mobile heterogeneous networks (HetNets), which combine various access network technologies, have become an imminent trend. As such, applying the THz technology to HetNets is a promising way to improve the transmission rate as well as the capacity and achieve a throughput at the level of Tbps \cite{li2018secure}.  Despite the high path loss and highly directional antenna requirements, these disadvantages could change into satisfactory features while operating in the femtocell regime. The deployment of femtocells reduces the required distance between both the active base-station and the user, while maintaining  high signal to interference and noise ratio (SINR) at the receiver.
Through such setup,  femtocell base-stations  improve the principle of frequency reuse and increase the capacity of the THz band systems. These access points are applied as  portals to in-home service and automation,  metro-stations, shopping malls,  traffic lights and many other applications. As such, a novel era of communications via THz signals for mobile HetNets will be witnessed through the installation of these access points. Based on  several metrics including the environment, the quality  and type of communication service,  both picocells  and  femtocells will be  accordingly collocated  within  the  macrocell footprint, as illustrated in Fig. \ref{fig:hetnet}. In fact, the authors in \cite{giordani2019towards} note that 6G technology will allow cell-less architectures and compact integration of multiple frequencies  and  communication  technologies. Such vision may be achieved by deploying multiple connectivity approaches and providing support for diverse and heterogeneous radios in the devices. Both seamless mobility support without overhead from handovers  and   QoS guarantees even in challenging mobility scenarios  will be assured via the cell-less network procedures. In addition, since ultra-dense (UD)-HetNets are bound to networks of big data, the authors in \cite{li2018ultra} introduced an AI-based network framework for energy-efficient operations.  The presented  framework supplies the network with the abilities of learning and inferring by analyzing the collected big data and then saving energy from both large scales (base station operation) and small scales (proactive caching and interference-aware resource allocation). The fact that THz communication is composed of access points in pervasive WiFi networks or base-station clustering in heterogenous networks,  reinforcement learning may be deployed. Such  self-organization capability is needed in THz communication to allow femtocells to autonomously recognize available spectrum  and adjust their parameters subsequently. These cells will therefore operate under restrictions of avoiding intra/inter-tier interference and satisfy QoS requirements \cite{6994301}.
  
\subsection{Terahertz 3D Beamforming Technology}
 One of the anticipated key enablers of THz wireless systems is 3D MIMO technology. In fact, real-world channels emphasize 3D characteristics leaving 2D MIMO techniques suboptimum \cite{cheng2014communicating}. 3D beamforming emerges as a solution to allow the construction of directional beams, extend the communication range as well as lower the interference level. Such technology holds a lot of promise to mitigate the unavoidable  path loss experienced by the THz channel. In specific, the vertical beam pattern possesses a complete active correspondence per resource and per user equipment. 3D beamforming can also increase the  strength of the signal by allowing the  vertical main lobe to be located precisely at the receiver at any position. By adopting beam coordination or MIMO schemes, the alteration in vertical dimension has the potential to capitalize  on additional diversity or spatial separation. This will lead to increasing  the  quality of the signal or increasing the number of supported users \cite{6772147}.
The ability to control the array’s radiation pattern in 3D is nonetheless helpful to manipulate the multipath environment resulting in a constructive addition of the many signal components at the location of the expected receiver. On a similar frontier, the authors in \cite{orazbayev2016tunable} showcase tunable beam  steering  devices based   on   multilayer   graphene-dielectric   metamaterials. Since the effective refractive  index  of  such  metamaterials  can  be  altered by  changing the chemical potential of each graphene layer, the spatial distribution of the phase of the transmitted beam can be tailored. This results in establishing mechanisms  for  active  beam  steering resulting  tunable  transmitter/receiver  modules for imaging and sensing at THz frequencies.

In addition, in order to mitigate the severe Doppler effect in mmW/THz massive MIMO systems, the authors in \cite{you2017bdma} proposed a beam division multiple access technique with per-beam synchronization capability in time and frequency. The authors verified via simulations the effectiveness of the proposed technique,  where they showed that both the  channel delay spread and Doppler frequency spread can be decreased  via per-beam synchronization.
This results in reducing the overall system overhead and outperforming conventional techniques in typical mobility scenarios. 
\subsection{Terahertz Communication for Urban Environments}

In 2016, Facebook launched a new project called “Terragraph” to provide crowded urban areas with a high-speed internet service \cite{choubey2016introducing}. Terragraph adopted the mmW band, specifically the 60 GHz frequency range, and utilized distributed access points over the existing city infrastructure to allow quick, easy, low cost, and tractable installation. The multiple access points communicate with each other creating mesh network over the city instead of lying down optical fiber that is unfeasible in the high-density urban environments. The Terragraph introduced a powerful solution that uses 7-14 GHz bandwidth, which is considered the largest commercial radio band ever used till now. Moreover, it is a licensed free spectrum until this moment, which further decreases the mesh network deployment cost. Therefore, the Terragraph network introduced a good network connectivity solution to connect the service provider with end users via Gbps links using existing urban physical assets such as traffic light poles and lamps posts. 

\begin{figure}[h!] 
  \centering
  \includegraphics[width=3.35in]{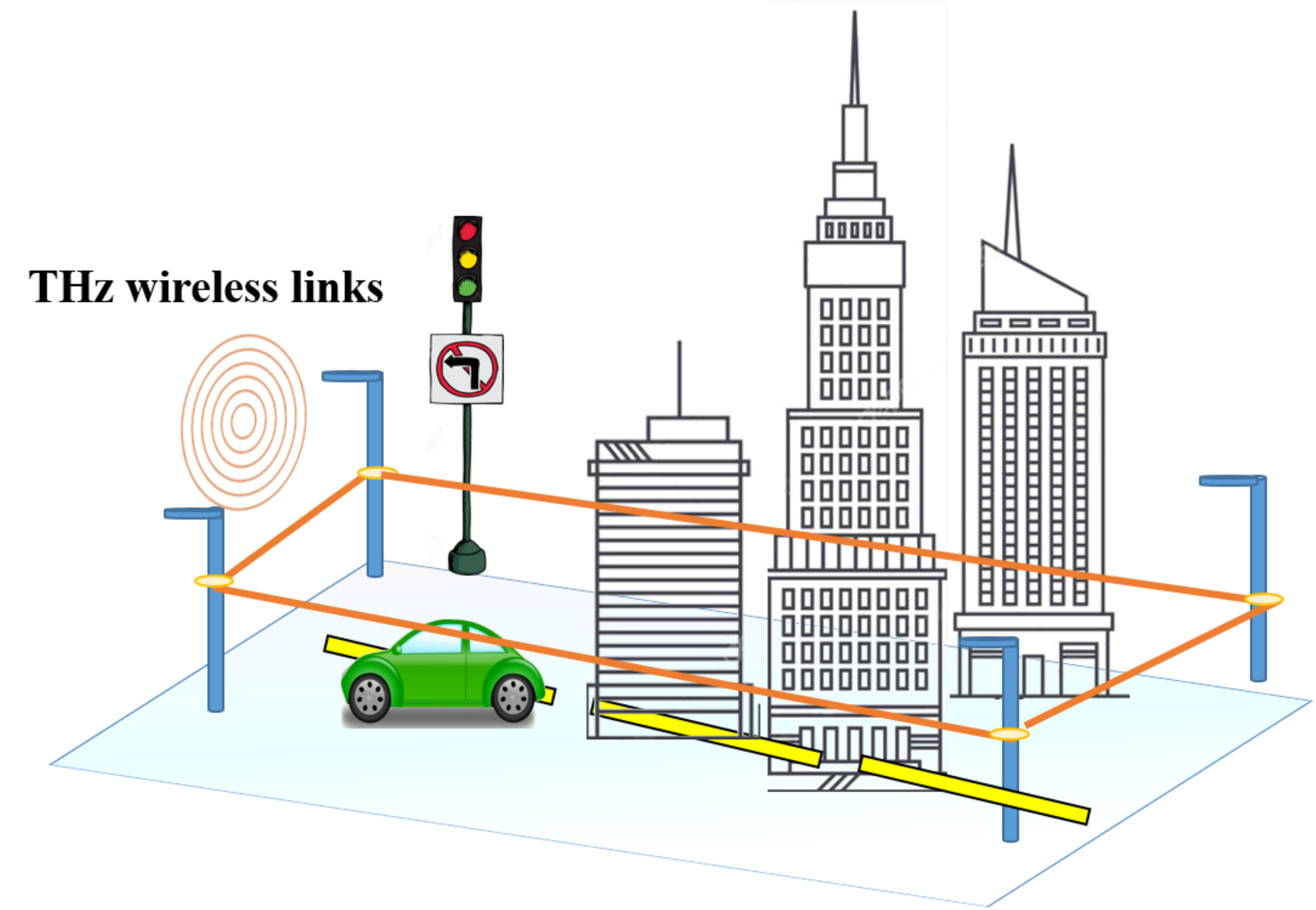}
  \caption{THz wireless links as candidates for establishing communication in an urban environment.}
  \label{fig:urban}
\end{figure}

Despite the advantages mentioned above of the wireless mesh network solution, several obstacles can limit its performance and affect using it for similar scenarios in the future. First, the mmW frequency bands for the International Mobile Communications (IMT) 2020 are still under study, where the decision is expected to be taken in the World Radio Conference (WRC) 2019 that will be held on Nov. 2019 \cite{WinNT1}. Second, the mmW band is expected to become crowded in the next decade. Thus, it will not be possible to accommodate more users and satisfy the exponential increase in population and data communications services. Finally, the mmW signal attenuates in the rain environment; thus the mesh network can be down under such circumstances. In other words, although the Terragraph project proposed rerouting techniques to avoid the scenario of link outage, rain can put most of the network in a blackout. As such, the THz frequency band provide a reliable wireless network access alternative with multiple backup links to avoid outages especially that it can work under different weather conditions. The THz band shall accommodate future population increase, urban environment rapid changes and new hungry rate services.
An illustration of THz communication  for urban environment is demonstrated in Fig. \ref{fig:urban}.
\subsection{Terahertz Automotive Applications}
\subsubsection{Vehicle to Infrastructure Communication}
\begin{figure}[h!] 
  \centering
  \includegraphics[width=3.35in]{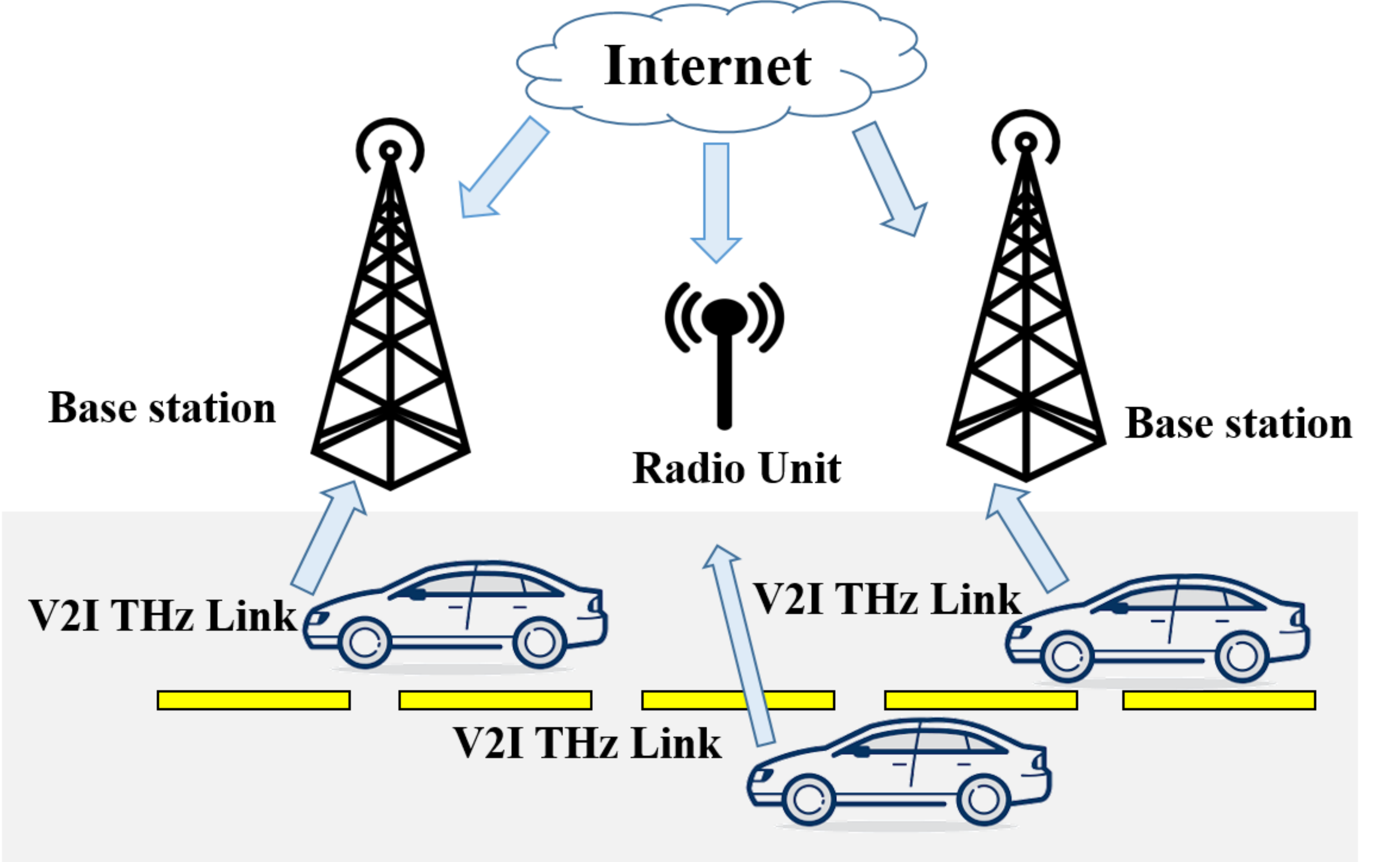}
  \caption{Envisioned V2I future communication scenarios utilizing the THz frequency band.}
  \label{fig:V2I}
\end{figure}

 The progress witnessed in the vehicle to infrastructure communication is considered a major milestone  in the automotive industry. The initiation of a  communication link that connects wireless between  vehicles and the road-side infrastructures paves the way towards the the deployment of fully autonomous and smart transportation systems. According to the literature \cite{araniti2013lte}, the Long Term Evolution (LTE) has been the standard wireless interface which supports communications in vehicular environments. However, due to the stringent requirements of the users and  the demands of the market in terms of higher data rates and lower latency to mobile users, new solutions must arise to fulfill the needs of next-generation networks. As such, the authors in \cite{giordani2019lte} discussed the feasibility of establishing vehicle to infrastructure communications using higher frequencies, namely the mmW, to support automotive applications.
Despite the anticipated benefits associated with mmW technology in both metropolitan and mobile highway scenarios, a number of challenges still arise. These include path-loss, shadowing, high directionality of beams as well as high sensitivity to blockage. Thereby, the THz frequency band seems to be a better alternative  especially due to its capability of supporting  the required estimated throughput of terabyte per driving hour \cite{lu2014connected,giordani2019lte,petrov2019unified}. A schematic diagram mimicking V2I communication using THz links is provided in Fig. \ref{fig:V2I}. As such, the high data rate communication, high-resolution radar sensing capabilities as well as the directional beam alignment capability of the THz transmitter and receiver result in such technology being  a stronger candidate for smart vehicular communication scenarios. 

Not only vehicle to infrastructure communication technology is evolving but also train to infrastructure (T2I) communication is developing towards smart rail mobility.  Indeed, since high-data rate  wireless connectivity with bandwidth beyond GHz is needed in  order to establish T2I and interwagon scenarios, the authors in \cite{guan2019measurement} demonstrated a complete study concerning measurement, simulation, and characterization of the T2I channel using the THz frequency band. Despite the high path loss of THz signals as well as the high mobility experienced by such high speed trains, the authors note that a robust THz link between the access points of the network can still be achieved. This is due to the fact that the user’s desired content may be distributed into several segments that are delivered individually to broadcast points based on the train’s schedule. Such procedure is facilitated by utilizing a proactive content caching scheme \cite{kanai2016proactive}, paving the way towards seamless data transmission. 

\subsubsection{Unmanned autonomous vehicles (UAVs)}
Unmanned autonomous vehicles (UAVs) have recently become accessible to the public. This resulted in several applications targeting both civilian and commercial domains. Typical examples involve weather monitoring, forest fire detection, traffic control, cargo transport, emergency search and rescue as well as communication relaying \cite{zeng2016wireless}. To deploy these applications, UAVs need to have a reliable communication link accessible at all times. For heights above  16 km, the effect of moisture is trivial; thus, THz attenuation is negligible. As such, THz can become a strong candidate to initiate reliable communications for varying UAV application scenarios.

 In comparison to free space optical, the THz frequency band is a sufficient technology since it will not only enable high-capacity UAV-UAV wireless backhaul but also allow a better substitute in alleviating the high mobility environment of UAVs. 
In fact, as a result of mobility,  communication links which suffer from the Doppler effect are minimized as carrier frequencies increase. Therefore, THz communication can establish high-speed communication links between two potentially dynamic locations \cite{mumtaz2017terahertz}.  UAVs also need short-distance secure  links  to receive instructions or transmit data before dispersing to fulfill their remote controlled or autonomous missions.  THz links are thereby considered a reliable venue for exchanging safety-critical information between UAVs as well as between the UAV and ground control stations. The large channel bandwidth of THz systems allows for specific protection measures against various standoff attacks like jamming and have the ability to completely hide information exchange.  Furthermore, THz links could be also utilized between UAVs and airplanes in order to support internet for  flights instead of using the satellite service. In this way, the UAV will act as a switchboard in the sky serving as an intermediary between the ground station and the airplane. 


\subsection{Terahertz Security Measures}
Despite the prevailing expectation of enhanced security for wireless data links operating at high-frequencies, the authors in \cite{ma2018security} show that an eavesdropper can intercept signals in LoS transmissions even when transmission occurs at high frequencies with narrow beams. The techniques the eavesdropper uses at high frequencies varies in comparison to those used for lower frequency transmissions.  For high frequencies, an object is placed in the path of the transmission to scatter radiation towards the eavesdropper. Hence, the authors  present a technique to mitigate such eavesdropping approach, which suggests characterizing the backscatter of the channel. If the signals incoming towards the transmitter can be measured and differentiated  from the variable backscattered off mobile objects or the environment, then  a sign of a probable attack would be through noticing any change, either an increase or a decrease, in the signal. Such technique provides an extra level of security especially when added to conventional counter-measures. Thus, to embed security into a directional wireless link, systems will necessitate original physical layer components and protocols for channel estimation. The presented work implies the significance of physical layer security in THz wireless networks and the urge for transceiver designs that include new counter-measures.

\section{Conclusion}
\label{Sec8}
To satisfy the demands for higher data rates and  support services of various traffic patterns, novel and efficient wireless technologies for a range of transmission links ought to be developed. As 5G networks are being deployed in various parts across the globe utilizing the mmW frequencies, the research community is exploring the THz frequency band as a revolutionary solution to support beyond 5G networks and enable applications that couldnt be deployed through 5G due to unforeseen difficulties. In this paper, a comprehensive survey has been presented for THz wireless communication in an attempt to review the devices, channel models as well as applications associated with the development of THz system architectures. As such, the THz frequency  generation techniques have been extensively reviewed, where the progress in electronics, photonics as well as plasmonics techniques has been highlighted. Moreover, the THz channel models which capture the channel characteristics and propagation phenomena have been presented for different use-case scenarios. An extensive  comparison was further   conducted to point the differences between  THz wireless and other existing technologies including  mmW, infrared, visible light and ultraviolet communication indicating  the foreseen potential upon the deployment of the THz band. In addition, a plethora of applications which tackle nano, micro as well as macro-scale THz scenarios have been demonstrated. Further, the standardization activities as well as  the investigation efforts of frequency bands up to 3000 GHz are demonstrated  indicating the collaborative efforts bringing THz science communities together. Finally, a number  of promising techniques and  deployment opportunities are presented in an attempt to efficiently satisfy the needs of future networks and face the technical challenges associated with implementing THz communication. Actually,  with the  continuous progress  in  THz  devices,  new foundations for  rapid  development of  practical   systems will be established. With the emergence of THz  communication  systems, societies will be expecting  near-instant, unlimited wireless connectivity with capabilities extending beyond the 5G networks. Virtual reality, HD streaming as well as automation for the internet of things  are amongst the many promising applications that shall be brought through the THz frequency band.

 \bibliographystyle{IEEEtran}

\end{document}